\documentclass[fleqn,usenatbib]{mnras}

\usepackage{newtxtext,newtxmath}
\usepackage[T1]{fontenc}
\usepackage{ae,aecompl}

\usepackage{graphicx}	
\usepackage{amsmath}	
\usepackage{subfig}
\usepackage{amsmath, epsfig,natbib}
\usepackage{multicol}
\usepackage{color,ulem}
\usepackage{newtxtext,newtxmath}
\definecolor{webgreen}{rgb}{0,.5,0}
\definecolor{webbrown}{rgb}{.6,0,0}
   
 \newcommand{\kms}{\mbox{$\>{\rm km\, s^{-1}}$}}
\newcommand{\pc}{\>{\rm pc}}
\newcommand{\kpc}{\mbox{$\>{\rm kpc}$}} 

\newcommand{\Gyr}{\mbox{$\>{\rm Gyr}$}}
\newcommand{\Myr}{\mbox{$\>{\rm Myr}$}}

\newcommand{\Msun}{\>{\rm M_{\odot}}}

\newcommand\degrees{^\circ}
\newcommand{\avg}[1]{\mbox{$\left<{#1}\right>$}}

\title [Lopsidedness in minor mergers] 
{Genesis of morpho-kinematic lopsidedness in minor merger of galaxies}

\author[S. Ghosh et al.]
	{Soumavo Ghosh$^{1,2}$\thanks{E-mail : ghosh@mpia-hd.mpg.de},
	      Kanak Saha$^{2}$,
		  Chanda J. Jog $^{3}$,
		  Francoise Combes$^{4}$,
	Paola Di Matteo$^{5}$\\
$^{1}$ Max-Planck-Institut f\"{u}r Astronomie, K\"{o}nigstuhl 17, D-69117 Heidelberg, Germany\\
$^2$  Inter-University Centre for Astronomy and Astrophysics, Pune 411007, India\\
$^3$  Department of Physics, Indian Institute of Science, Bangalore 560012, India\\
$^4$  Observatoire de Paris, LERMA, College de France, CNRS, PSL University, Sorbonne University, Paris, France\\
$^5$  GEPI, Observatoire de Paris, PSL Research University, CNRS, Place Jules Janssen, 92190, Meudon, France\\
} 
\date{Accepted 2022 February 16. Received 2022 February 16; in original form 2021 May 21}

\pubyear{2021}

\begin{document}
\label{firstpage}
\pagerange{\pageref{firstpage}--\pageref{lastpage}}
\maketitle


\begin{abstract} 

An $m=1$ lopsided asymmetry is common in disc galaxies. Here, we investigate the excitation of an $m=1$ lopsidedness in host galaxies during minor mergers while choosing a set of 1:10 merger models (with varying orbital configurations, morphology of the host galaxy) from the GalMer galaxy merger library. We show that a minor merger triggers a prominent $m=1$ lopsidedness in stars of the host galaxy. The strength of the $m=1$ lopsidedness undergoes a transient amplification phase after each pericenter passage of the satellite, in concordance with past findings of exciting an $m=1$ lopsidedness by tidal encounters. However, once the merger happens, and the post-merger remnant readjusts itself, the lopsidedness disappears in short time-scale ($\sim$ 500-850 $\Myr$). Furthermore, a delayed merger can drive a prolonged ($\sim$2 $\Gyr$) lopsidedness in the host galaxy. We demonstrate that the $m=1$ lopsidedness rotates with a well-defined pattern speed which is much slower than the $m=2$ bar pattern speed, and is retrograde with respect to the bar. This gives rise to a dynamical scenario where the Inner Lindblad resonance of the $m=1$ lopsidedness falls in between the corotation and the Outer Lindblad resonance of the $m=2$ bar mode. A kinematic lopsidedness also arises in the host galaxy; the resulting temporal variation closely follows that of the density lopsidedness. The minor merger also triggers a transient off-centred stellar disc-dark matter halo configuration due to the tidal encounter with the satellite.

\end{abstract}


\begin{keywords}
{galaxies: evolution - galaxies: formation - galaxies: haloes - galaxies: kinematics and dynamics - galaxies: structure}
\end{keywords}


\section{Introduction}
\label{sec:intro}

Past observational studies revealed that disc galaxies often exhibit an $m=1$ distortion or  lopsidedness in the outskirts of the disc. Lopsidedness is common in the spatial distribution of the neutral hydrogen (H~{\sc i}) which extends further out than the stellar disc \citep[e.g., see][]{Baldwinetal1980,RichterandSancisi1994,Haynesetal1998,Matthewsetal1998,Angirasetal2006,vanEymeren2011} as well as in the spatial distribution of stars \citep[e.g., see][]{Blocketal1994,RixandZaritsky1995,Bournaudetal2005,Reichardetal2008,Zaritskyetal2013}. Previous work by \citet{KalberlaandDedes2008} showed the presence of a lopsidedness in the H~{\sc i} distribution of the Milky Way whereas a recent work by \citet{Romeroetal2019} suggested a lopsided (warped) stellar disc for the Milky Way. Simultaneous occurrence of an $m=1$ lopsided distortion and the $m=2$ bar and spiral arms are also common \citep[e.g., see][]{RixandZaritsky1995, Bournaudetal2005, Zaritskyetal2013}. The magnitude of the $m=1$ lopsidedness is shown to correlate with the strength of the spiral arms, but is not correlated with the occurrence of the bar \citep[e.g., see][but also see \citealt{Bournaudetal2005}]{Zaritskyetal2013}. Signature of lopsidedness has been reported in the H~{\sc i} velocity fields of galaxies as well \citep[e.g. ][]{Swatersetal1999,Schoenmakersetal1997,vanEymerenetal2011}. A lopsided pattern in the density distribution can give rise to a kinematic lopsided feature \citep[e.g., ][]{Jog1997,Jog2002}. Indeed, such a co-existence of morphological and kinematic lopsidedness has been shown observationally in a sample of galaxies from the WHISP (Westerbork H~{\sc i} Survey of Spiral and Irregular Galaxies) survey \citep[see][]{vanEymerenetal2011,vanEymeren2011}.
\par
A variety of physical mechanisms has been identified which can excite an $m=1$ lopsided pattern in a disc galaxy. For example, the disc response to halo lopsidedness arising due to tidal interactions \citep{Jog1997} or merging of a satellite galaxy \citep{ZaritskyandRix1997} or a tidal encounter \citep{Bournaudetal2005,Mapellietal2008}, and asymmetric gas accretion \citep{Bournaudetal2005} can lead to an excitation of an $m=1$ lopsidedness. For a detailed exposition of this field, see the review in \citet{JogandCombes2009}. Also, an off-set disc in a spherical dark matter (hereafter DM) halo can excite a lopsidedness feature \citep{Noordermeer2001,PrasadandJog2017}.  A recent study by \citet{SahaandJog2014} showed that a leading $m=1$ lopsidedness can take part in the outward angular momentum transport, thus facilitating the inflow of gas from the outer regions of galaxy. However, little is known about the pattern speed of the $m=1$  lopsidedness. Observationally, the pattern speed of lopsidedness has not been measured till date. Earlier theoretical works \citep[e.g., see][]{RixandZaritsky1995,Jog1997} have assumed a null pattern speed, for simplicity. Further theoretical explorations revealed that the \textit{slowly} varying global $m=1$ modes can survive for longer times in the near-Keplerian central regions of M~31 \citep{Tremaine2001} as well as in the pure exponential discs in spiral galaxies \citep{Sahaetal2007}.  Previous works by \citet{JunqueiraCombes1996,Baconetal2001} also measured the pattern speed of an $m=1$ lopsidedness in the central regions ($\sim$ few tens of pc) of M~31-like galaxy models using numerical simulations. Measuring the pattern speed of the lopsided asymmetry is extremely crucial as it can potentially shed light about the dynamical role of the lopsidedness in the secular evolution and the angular momentum transport. Furthermore, it can provide important clues about the generating mechanisms of the lopsidedness  \citep[e.g., see discussion in][]{Jog2011}.
\par
Also, a few studies of mass modelling from the rotation curve have furnished evidences/indications that there could be an off-set (ranging between $\sim 1-2.5 \kpc$) between the baryonic disc and the DM halo, for example, in NGC~5055 \citep{Battagliaetal2006}, in one galaxy residing in the galaxy cluster Abell 3827 \citep{Masseyetal2015}, and also in M~99 \citep{Cheminetal2016}. Furthermore, a theoretical study by \citet{Kuhlenetal2013} reported an off-set of $300-400 \pc$ between the density peaks of the baryonic disc and the DM halo in a Milky Way-like galaxy from the high-resolution cosmological hydrodynamics {\sc{Eris}}. This off-set is seen to be long-lived. An off-centred nucleus can result in an unsettled central region \citep[e.g.,][]{MillerandSmith1992}. Indeed, such a sloshing pattern in the central regions has been reported in a sample of remnants of advanced mergers of galaxies \citep{JogandMaybhate2006}. In the past, several theoretical efforts focused on studying the disc-DM halo response to an interaction with a passing-by satellite or a unbound encounter, by means of linear perturbation theory or by numerical simulations. It was shown that such interactions lead to the excitation of  coherent modes in the DM halo distribution. It can also lead to the production of strong disturbances in the baryonic disc, leading to excitation of a vertical $m=1$ warp mode and the lopsidedness \citep[e. g., see][]{Weinberg1989,Weinberg1994,Weinberg1998,VesperiniandWeinberg2000,Choi2007,Gomezetal2016,Laporteetal2018}. Furthermore, \citet{GaoandWhite2006} studied the off-set in the dark matter halo distribution from a high-resolution cosmological simulation, and showed that only $\sim 7$ percent of the Milky Way-like dark matter haloes display an off-set of more than $20$ percent between their halo density centre and the barycentre.
\par
On the other hand, minor merger of galaxies is common in the hierarchical formation scenario of galaxies \citep{Frenketal1988,CarlbergandCouchman1989,LaceyandCole1993,Jogeeetal2009,Kavirajetal2009,FakhouriandMa2008}. This mechanism has a number of dynamical impacts on the kinematics as well as on the secular evolution of galaxies, such as disc heating and the vertical thickening of discs \citep{Quinnetal1993,Walkeretal1996,VelazquezandWhite1999,Fontetal2001,Kazantzidisetal2008,Quetal2011a}, slowing down the stellar discs of the post-merger remnants \citep{Quetal2010,Quetal2011b}, enhancing star formation \citep[e.g., see][]{Kaviraj2014}, transferring angular momentum to the dark matter halo via action of stellar bars \citep{Debattistaetal2006,SellwoodandDebattista2006}, and weakening of the stellar bars in the post-merger remnants \citep{Ghoshetal2020}. Furthermore, a recent numerical study by \citet{Pardyetal2016} has shown that a dwarf-dwarf merger can produce an off-set bar and a highly asymmetric stellar disc that survives for $\sim 2 \Gyr$. This serves as a plausible explanation for the off-set bar \citep{Kruketal2017} found in many Magellanic-type galaxies. While a minor merger can excite lopsidedness in disc galaxies \citep[e.g.,][]{Bournaudetal2005,Mapellietal2008}, the exact role of different orbital parameters, Hubble type of the companion, remain unexplored in the context of excitation of an $m=1$ lopsidedness during a minor merger event. 
\par
In this paper, we systematically investigate the generation of an $m=1$ lopsidedness in the density and the velocity fields of the host galaxy in a minor merger scenario while varying different orbital parameters, nature of the host galaxies. Also, we study in details whether a minor merger of galaxies can produce an off-set between the baryonic and the DM halo density distribution.  For this, we make use of the publicly available GalMer library \citep{Chillingarianetal2010} which offers the study of the physical effects of minor merger of galaxies, encompassing a wide range of cosmologically motivated initial conditions. Thus, this database is appropriate for fulfilling the goal of this paper. 
\par
The rest of the paper is organised as follows. Section~\ref{sec:simu_setup} provides a brief description of the GalMer database and the minor merger models used here. Section~\ref{sec:diskhalooffset} provides the details of the disc-DM halo off-set configuration arising in minor merger models. Section~\ref{sec:Lopsidedness} quantifies the $m=1$ lopsided distortions present in the stellar disc of the host galaxy whereas section~\ref{sec:pattern_speed} provides the pattern speed measurement and the location of resonance points associated with the $m=1$ lopsidedness. Section~\ref{sec:flapping_mode} presents the details of the kinematic lopsidedness in the minor merger models. Section~\ref{sec:comparison_previousWork} compares the properties of the $m=1$ lopsidedness, as presented here, with the past literature. Sections~\ref{sec:discussion} and \ref{sec:conclusion} contain discussion and  the main findings of this work, respectively.

\section{Minor merger models -- GalMer database}
\label{sec:simu_setup}

\begin{table*}
\centering
\caption{Key parameters for the selected minor merger models from GalMer library.}
\begin{tabular}{ccccccccccc}
\hline
model$^{(1)}$ & $r_{\rm ini}$$^{(2)}$ & $v_{\rm ini}$$^{(3)}$ & $L_{\rm ini}$$^{(4)}$ & $E_{\rm ini}$$^{(5)}$ &spin$^{(6)}$ &  pericenter$^{(7)}$ & $T_{1,\rm  peri}$$^{(8)}$ & $T_{2, \rm peri}$$^{(9)}$ & $T_{\rm merger}$$^{(10)}$ & $T_{\rm end}$$^{(11)}$\\
& (kpc) & ($\times 10^2 \kms$) & ($\times 10^2 \kms \kpc$) & ($\times 10^4$ km$^2$ s$^{-2}$) &&  dist. (kpc) & (Gyr) & (Gyr) & (Gyr) & (Gyr)\\
\hline
gSadE001dir33 & 100 & 1.48 & 29.66 & 0. & up & 8.  & 0.5 & 1.1 & 1.55 & 3.8\\
gSadE001ret33 & 100 & 1.48 & 29.66 & 0. & down &8. &  0.5 & 1.3 & 1.95 & 3.8 \\
gSadE002dir33 & 100 & 1.52 & 29.69 & 0.05 & up & 8. & 0.45 & 1.2 & 1.55 & 3. \\
gSadE002ret33 & 100 & 1.52 & 29.69 & 0.05 & down & 8. & 0.45 & 1.4 & 2. & 3. \\
gSadE003dir33 & 100 & 1.55 & 29.72 & 0.1 & up & 8. & 0.45 & 1.25 & 1.95 & 3. \\
gSadE003ret33 & 100 & 1.55 & 29.72 & 0.1 & down & 8. & 0.45 & 1.5 & 2.25 & 3. \\
gSadE004dir33 & 100 & 1.48 & 36.33 & 0. & up & 8. & 0.5 & 1.2 & 1.7 & 3. \\
gSadE004ret33 & 100 & 1.48 & 36.33 & 0. & down & 8.  & 0.5 & 1.75 & 2.85 & 3. \\
gSadE005dir33 & 100 & 1.52 & 36.38 & 0.05 & up & 16. & 0.5 & 1.35 & 1.85	& 3. \\
gSadE005ret33 & 100 & 1.52 & 36.38 & 0.05 & down & 16. & 0.5 & 1.85 & 2.7 & 3. \\
gSadE006dir33 & 100 & 1.55 & 36.43 & 0.1 & up &  16. & 0.45 & 1.45 & 2. & 3. \\
gSadE006ret33 & 100 & 1.55 & 36.43 & 0.1 & down & 16. & 0.45 & 1.95 & 2.85 & 3. \\
gS0dE001dir33 & 100 & 1.48 & 29.66 & 0. & up & 8. & 0.5 & 1.2 & 1.55 & 3.8 \\
gS0dE001ret33 & 100 & 1.48 & 29.66 & 0. & down &  8. & 0.55 & 1.7 & 2.2 & 3.8 \\
gSbdE001dir33 & 100 & 1.48 & 29.66 & 0. & up & 8. & 0.5 & 1.05 & 1.35 & 3 \\
gSbdE001ret33 & 100 & 1.48 & 29.66 & 0. & down & 8. & 0.5 & 1.35 & 1.85 & 3 \\
\hline
\end{tabular}
\centering
{ (1) GalMer minor merger model; (2) initial separation between two galaxies; (3)  absolute value of initial relative velocity; (4)  $L_{\rm ini} = |{\bf r}_{\rm ini} \times {\bf v}_{\rm ini}$|}; (5) $E_{\rm ini} = \frac{1}{2}v_{\rm ini}^2 -G(m_1+m_2)/r_{\rm ini}$, with $m_1 = 2.3 \times 10^{11} \Msun$, and  $m_2 = 2.3 \times 10^{10} \Msun$; (6) orbital spin; (7) pericenter distance; (8) epoch of first pericenter passage; (9)  epoch of second pericenter passage; (10) epoch of merger; (11) total simulation run time. Columns (2)-(7) are taken from \citet{Chillingarianetal2010}.
\label{table:key_param}
\end{table*}

The publicly available GalMer \footnote{ available on \href{http:/ /galmer.obspm.fr} {http://galmer.obspm.fr}} library offers a suite of $N$-body+smooth particle hydrodynamics (SPH) simulations of  galaxy mergers that can be used to probe the details of galaxy formation through hierarchical merger process. It offers three different galaxy interaction/merger scenarios with varying mass ratio -- the 1:1 mass ratio mergers (giant-giant major merger), 1:2 mass ratio mergers (giant-intermediate merger), and 1:10 mass ratio mergers ( giant-dwarf minor merger). An individual galaxy model is comprised of a non-rotating spherical dark matter halo, a stellar and a gaseous disc (optional), and a central non-rotating bulge (optional). The bulge (if present) and the dark matter halo are modelled using a Plummer sphere \citep{Plummer1911} and the baryonic discs (stellar, gaseous) are represented by the Miyamoto-Nagai density profiles \citep{Miyamoto-Nagai1975}. The mass of the stellar disc varies from gS0, gSa-type ($9.2 \times 10^{10} \Msun$) to late-type gSd models ($5.8 \times 10^{10} \Msun$). Similarly, the bulge mass also decreases from gS0, gSa-type ($2.3 \times 10^{10} \Msun$) to late-type gSd models ($0$, no bulge). For details of the other structural parameters, the reader is referred to \citet[see Table.~1 there]{Chillingarianetal2010}. The total number of particles ($N_{tot}$) varies from a giant-dwarf interaction ($N_{tot}$ = 480, 000) to a giant-giant interaction ($N_{tot}$ = 120, 000). Similarly, the number of particles assigned to each of the sub-components (e.g., disc, bulge, DM halo) varies with the Hubble type of the galaxy. For example, the number of stellar particles ($N_{\rm star}$) varies from $3.2 \times 10^5$ for the gS0-type to $1.6 \times 10^5$ for the gSb-type model, with $20$ percent of $N_{\rm star}$ are assigned to model the bulge and the rest $80$ percent of the particles are assigned to model the stellar disc in each case. The number of particles used to model the DM halo ($N_{\rm DM}$) is $1.6 \times 10^5$, and is kept fixed for all Hubble types of host galaxies \citep[for details see Table~6 in][]{Chillingarianetal2010}. For a 1:10 mass ratio merger (giant-dwarf minor merger), a dwarf galaxy is a re-scaled version of a giant host galaxy (of same Hubble type). The only difference for the satellite is the total mass, and the number of particles used, which are ten times lower than those of the giant galaxy, and the size of the satellite becomes $\sqrt{10}$ times smaller than the host giant galaxy. To illustrate further, for a giant S0-type (gS0) galaxy, $N_{\rm star} = 3.2 \times 10^5$ and $N_{\rm DM} = 1.6 \times 10^5$ whereas for a dwarf S0-type (dS0) galaxy, $N_{\rm star} = 3.2 \times 10^4$ and $N_{\rm DM} = 1.6 \times 10^4$  \citep[for details see Table~6 in][]{Chillingarianetal2010}.
\par
Following \citet{MihosandHernquist1994}, a `hybrid particle' scheme is implemented to represent the gas particles in these simulations. In this prescription, they are characterised by two masses, namely, the gravitational mass ($M_i$) which is kept fixed during the entire simulation run, and the gas mass ($M_{i, gas}$), changing with time, denoting the gas content of the particles \citep[for details see][]{Dimatteoetal2007,Chillingarianetal2010}. Gravitational forces are always calculated using the gravitational mass, $M_i$, while the hydrodynamical quantities make use of the time-varying gas mass, $M_{i, gas}$. The gas fraction in a galaxy model increases monotonically from the early Sa-type galaxies (10 per cent of the stellar mass) to the late Sd-type galaxy (30 per cent of the stellar mass).  A suitable empirical relation is adopted in the simulations to follow the star formation process which reproduces the observed Kennicutt-Schmidt law for the interacting galaxies. The simulation models also include recipes for the gas phase metallicity evolution as well the supernova feedback. 
\par
The simulations are run using a TreeSPH code by \citet{SemelinandCombes2002}. For calculating the gravitational force, a hierarchical tree method \citep{BarnesandHut1986} with a tolerance parameter $\theta = 0.7$ is employed which includes terms up to the quadrupole order in the multipole expansion. The gas evolution is achieved by means of smoothed particle hydrodynamics \citep[e.g.][]{Lucy1977}. A Plummer potential is used to soften gravitational forces. The adopted softening length ($\epsilon$) varies with different merger scenarios, with $\epsilon = 200 \pc$ for the giant-intermediate and giant-dwarf merger models and $\epsilon = 280 \pc$ for the giant-giant merger models. The equations of motion are integrated using a leapfrog algorithm with a fixed time step of $\Delta t = 5\times 10^5$ yr \citep{Dimatteoetal2007}. The galaxy models are first evolved in isolation for $1 \Gyr$ before the start of merger simulation \citep{Chillingarianetal2010}.
\par
For this work, we consider a set of giant-dwarf minor merger models with varying morphology for the host  galaxy (see Table~\ref{table:key_param}).  In GalMer simulations, the orientation of an individual galaxy in the orbital plane is described completely by the spherical coordinates, $i_1$, $i_2$, $\Phi_1$, and $\Phi_2$ \citep[for details, see fig.~3 of ][]{Chillingarianetal2010}. However, the  GalMer library provides only one orbital configuration for the  giant-dwarf interaction, characterised by $i_1 =33^{\circ}$ and $i_2 =130^{\circ}$ \citep{Chillingarianetal2010}. We point out that this choice of inclination angle is in compliance with the expectation for a random distribution of inclinations between orbital planes and halo spins. Furthermore, using a high-resolution cosmological simulation,  \citet{KhochfarandBurkert2006} showed that the distribution of the angle between the orbital plane of the satellite and the spin plane of the halo  follows a sinus function; thus, justifying our choice of $33^{\circ}$ inclination \citep[for further details, see][]{Chillingarianetal2010}.
\par
For consistency, any merger model is referred as a unique string `{\sc [host galaxy][satellite galaxy][orbit ID][orbital spin]33}'. {\sc [host galaxy]} and {\sc [satellite galaxy]} denote the corresponding morphology types. {\sc [orbit ID]} denotes the orbit number as assigned in the GalMer library, and {\sc [orbital spin]} denotes the orbital spin vector (`dir' for direct and `ret' for retrograde orbits). The number `33' refers to $i_1 =33^{\circ}$ which is constant for all minor mergers considered here. The same sense of nomenclature is used throughout the paper, unless stated otherwise. 
\par
We define the epoch of merger, $T_{\rm mer}$, when the distance between the centre of mass of two galaxies becomes close to zero. Table.~\ref{table:key_param} contains the epoch of merger, $T_{\rm mer}$, for all minor merger models used for this work. Also, the epochs of the first and the second pericenter passages ($T_{1,\rm  peri}$,  $T_{2,\rm  peri}$) for these models are mentioned in Table~\ref{table:key_param}.

\section{Disc-DM halo off-set in minor mergers}
\label{sec:diskhalooffset}
%
\begin{figure*}
\centering
\includegraphics[width=\linewidth]{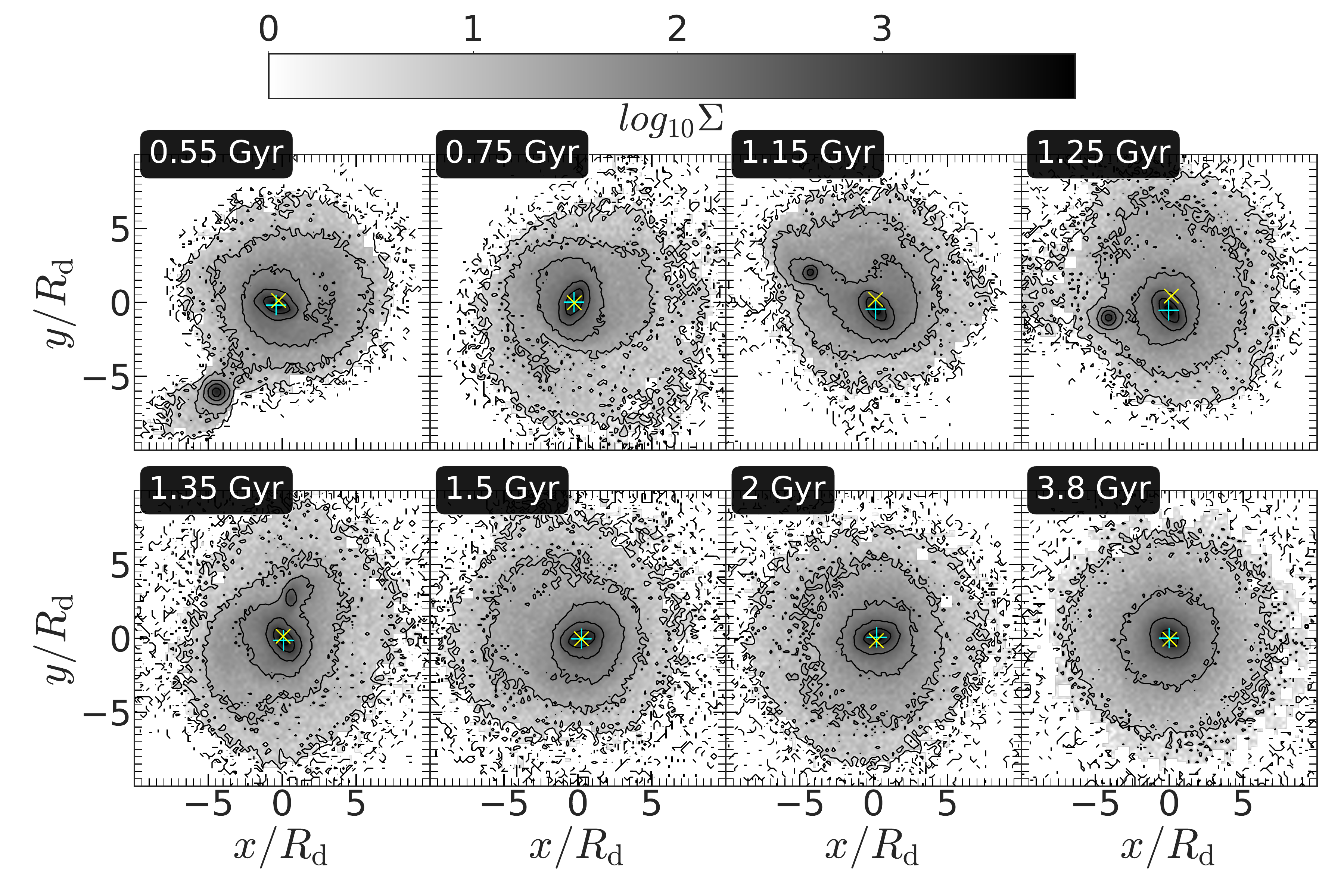}
\caption{Face-on stellar density distribution of host plus satellite (gSa+dE0) system, shown for the model gSadE001dir33 at different epochs before and after pericenter passages of the satellite galaxy.  Black lines denote the contours of constant surface density. The symbols `$+$' (in cyan) and `$\times$' (in yellow) denote the density-weighted centres for the stellar disc and the DM halo distributions of the host galaxy, respectively. Here, $R_{\rm d} = 3 \kpc$.}
\label{fig:density_illustration_collage}
\end{figure*}

\begin{figure*}
\centering
\includegraphics[width=\linewidth]{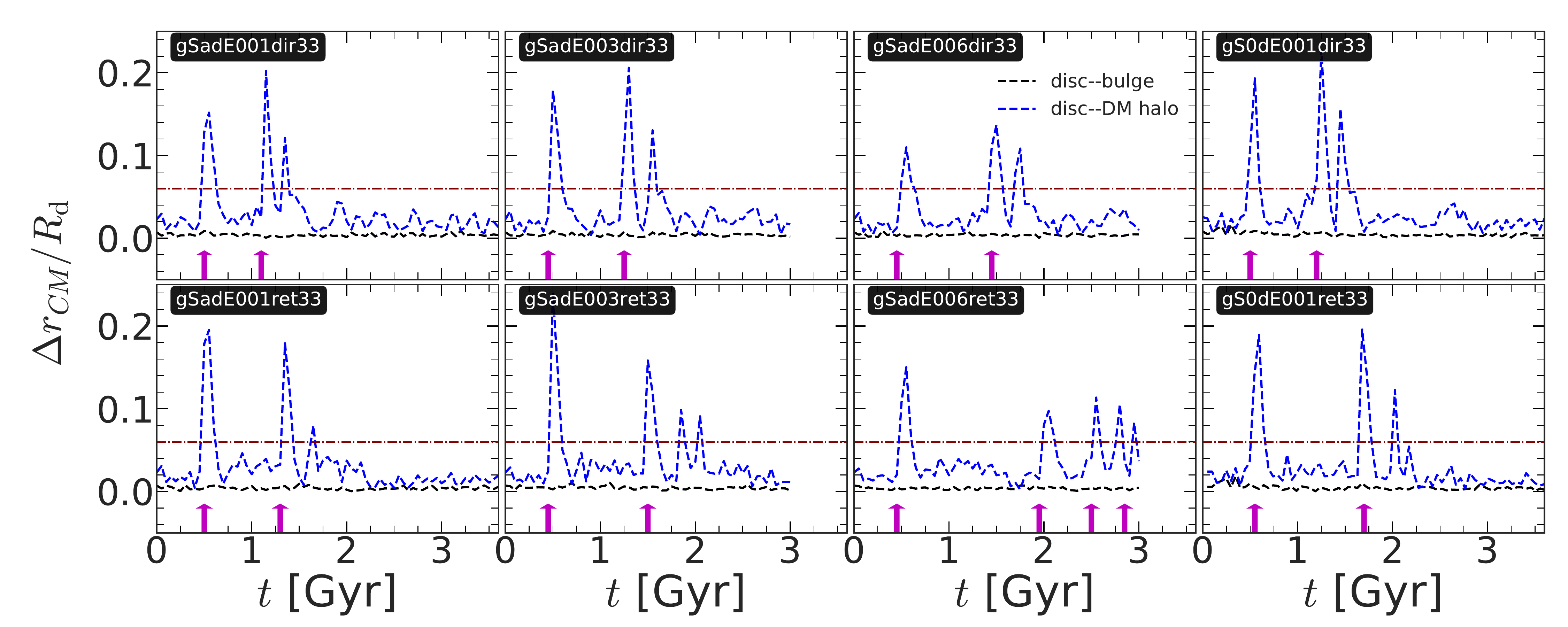}
\caption{ Disc-halo off-set in minor mergers : separation between the disc-bulge (black dashed line) and the disc-DM halo (blue dashed line) centres of the host galaxy, as a function of time are shown for different minor merger models. The barycentre of the host plus satellite galaxy system is used as the centre of reference. The vertical arrows (in magenta) denote the epochs of pericenter passages of the satellite galaxy. The horizontal dash line (in maroon) denotes the softening length ($\epsilon = 200 \pc$) adopted for the minor merger simulations. Here, $R_{\rm d} = 3 \kpc$.}
\label{fig:fig2}
\end{figure*}

\begin{figure*}
\centering
\includegraphics[width=1.\linewidth]{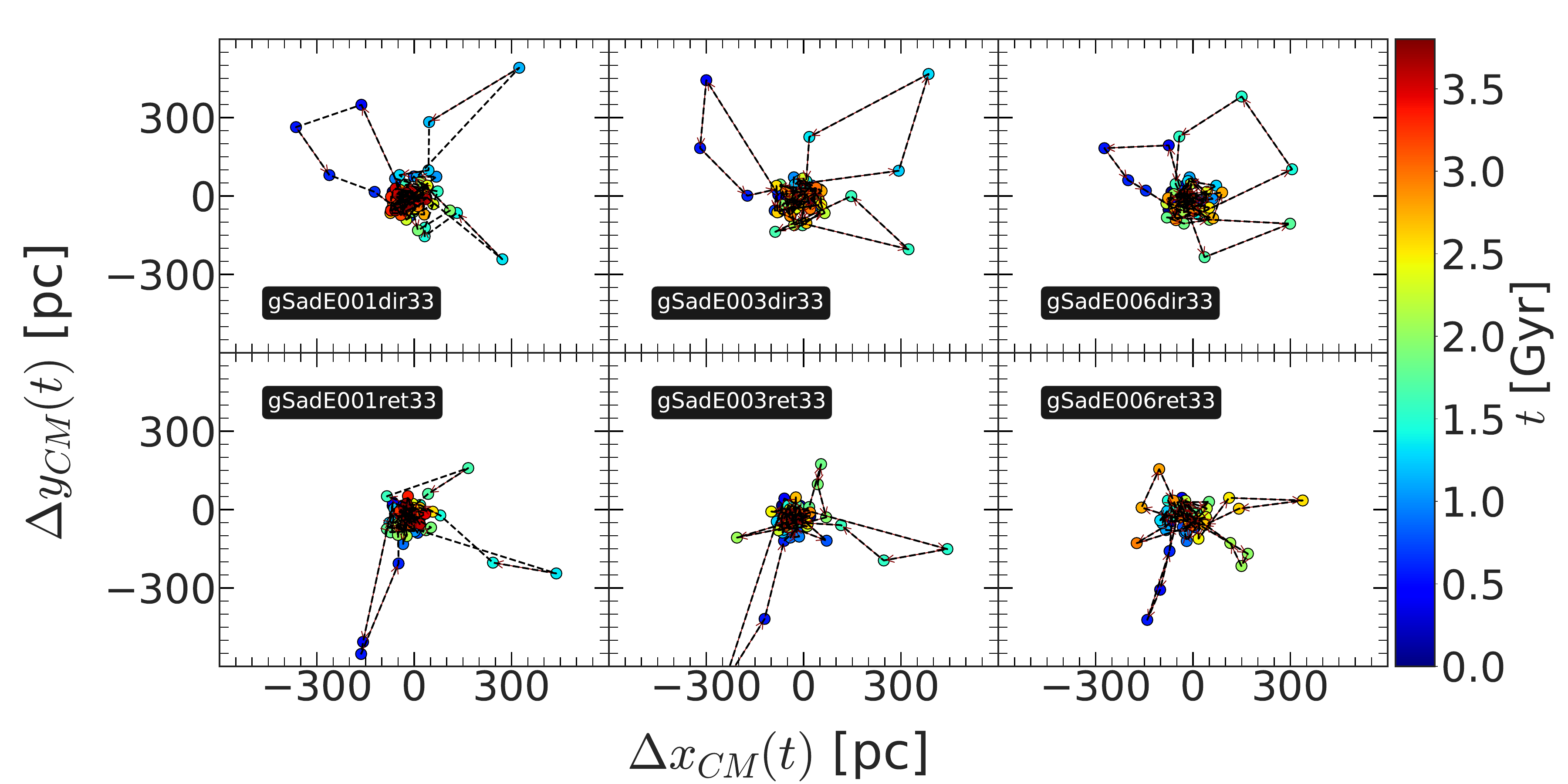}
\caption{Separation between the disc and DM halo centres of the host galaxy in the ($x-y$)-plane are shown for different minor merger models with varying orbital parameters. The barycentre of the host plus satellite galaxy system is used as the centre of reference. The colour bar denotes the epochs of the minor merger models.}
\label{fig:sloshing_of_centres}
\end{figure*}
First, we investigate whether a minor merger scenario can create a disc-DM halo off-set configuration in the models considered here. To achieve that, we first choose a minor merger model gSadE001dir33 from the GalMer library. Fig.~\ref{fig:density_illustration_collage} shows the corresponding face-on density distribution of the stellar particles from the host and the satellite galaxies (gSa+dE0) at different epochs, before and after pericenter passages. In this model, a giant Sa-type (gSa) galaxy experiences an interaction with a dwarf elliptical (dE0) galaxy. The satellite loses a part of its orbital angular momentum after each pericenter passage due to the dynamical friction and the tidal torque, thereby falling deep in the gravitational potential of the host galaxy and ultimately merges with it. After each pericenter passage of the satellite, the host galaxy displays a prominent distortion/asymmetry in the stellar density distribution (e.g., see $t = 0.75 \Gyr$ in the top panel of Fig.~\ref{fig:density_illustration_collage}).
\par
Next, we compute the density centres of the stellar disc and the DM halo of the host galaxy, and see whether they are concentric during and after the merger.We mention that, during the interaction, the host as well as the satellite galaxy form a tail-like feature due to the gravitational pull exerted on them. Therefore, the mass-weighted centre (centre-of-mass) could be misleading to locate the actual centre of the mass distribution. Consequently, one has to compute the density-weighted centres of the underlying mass distribution \citep[for detailed discussion, see][]{CasertanoandHut1985}. Following \citet{CasertanoandHut1985}, the density-weighted centre of a certain galactic component (here disc, and DM halo) is calculated using
\begin{equation}
{\bf{x}}_{d,j} = \frac{\sum_{i} {\bf x}_i \rho^{(i)}_j}{\sum_{i} \rho^{(i)}_j}\,,
\end{equation}
\noindent where  ${\bf x}_i$ is the three-dimensional position vector for the \textit{i}th particle, and $ \rho^{(i)}_j$ is the density estimator of order \textit{j} around the  \textit{i}th particle, and is evaluated as
\begin{equation}
\rho_j = \frac{j-1}{V(r_j)}m\,.
\end{equation}
\noindent Here, $m$ is the mass of the particle (equal for all particles of a certain galactic component), $r_j$ is the distance of the \textit{j}th particle from the particle around which local density is being estimated. Here, we choose $j=6$, as prescribed by \citet{CasertanoandHut1985}. Fig.~\ref{fig:density_illustration_collage} also shows the density-weighted centres of the stellar disc and the dark matter halo (indicated by  $+$ and $\times$, respectively) of the host galaxy at different times. Interestingly, the density-weighted centres of the stellar disc and the DM halo are seen to be separated by a finite amount after each pericenter passage of the satellite galaxy; thereby indicating the presence of an off-set between the stellar disc and the DM halo in the host galaxy. However, by the end of the simulation run ($t = 3.8 \Gyr$), these two centres are seen to coincide again.
\par
To investigate further, we calculate the density-weighted centres of stellar disc and the DM halo of the host galaxy as a function of time, for different minor merger models with varying orbital configuration, orbital energy (for details, see in Section~\ref{sec:simu_setup}). Here, we use the barycentre of the host plus satellite galaxy system as the centre of reference. 
Fig.~\ref{fig:fig2} shows the corresponding temporal evolution of the separation between the centres of the stellar disc and the DM halo of the host galaxy in different merger models. The separation/off-set ($\Delta r_{\rm CM} (t)$) at time $t$ is calculated as 
\begin{equation}
\Delta r_{\rm CM} (t) =\sqrt{\Delta x_{\rm CM}^2 (t)+\Delta y_{\rm CM}^2 (t)+\Delta z_{\rm CM}^2 (t)}\,,
\end{equation} 
\noindent where, $\Delta x_{\rm CM} (t) = x_d(t) -x_{dm} (t)$, $\Delta y_{\rm CM} (t) = y_d(t) -y_{dm} (t)$, $\Delta z_{\rm CM} (t) = z_d(t) -z_{dm} (t)$. Here ($x_d (t), y_d (t), z_d (t)$) and ($x_{dm} (t), y_{dm} (t), z_{dm} (t)$) denote the density-weighted centres of the stellar disc and the DM halo of the host galaxy at time $t$, respectively. {\footnote {After the merger happens, the stellar and the DM halo particles from the satellite get redistributed in the host galaxy. We check that, after the merger, the values of $\Delta r_{CM} (t)$ when calculated using particles from both the host and the satellite remains same as the $\Delta r_{CM} (t)$ values calculated by taking only the particles from the host galaxy.}} Fig.~\ref{fig:fig2} clearly demonstrates the presence of an off-set between the stellar disc and the DM halo of the host galaxy, after each time the host galaxy experiences a pericenter passage of the satellite galaxy. This off-set can be $\sim 400-600 \ \pc$ (2-3 times the softening length of the simulation), and is most prominent immediately after the pericenter passage. If there is sufficient time to adjust between two successive pericenter passages, the off-set decreases and goes below the softening length (and hence is not reliable). Once the satellite mergers, and the post-merger remnant gets some time ($\sim 250-400$ Myr) to readjust itself, this off-set disappears. In other words, the off-set between stellar disc and the DM halo in these minor merger models is a transient phenomenon. 
\par
Next, we compare how the generation of an off-set between the stellar disc and the DM halo of the host galaxy varies in models with different orbital configurations. We find that, each pericenter passage of the satellite triggers an off-set between stellar disc and the dark matter halo of the host galaxy in a generic fashion in these merger models. However, the actual value of such an off-set depends (weakly) on the distance of closest approach. For example, the model gSadE006dir33 has a pericenter distance of $16 \kpc$ as opposed to the model gSadE001dir33 which has a pericenter distance of $8 \kpc$  (for details, see Table~\ref{table:key_param}). The lower values of the off-set seen in the model gSadE006dir33 as compared to that of model gSadE001dir33 suggests a dependence on the distance of closest approach. Furthermore, the merger of the satellite with the host galaxy happens at a very late epoch for the model gSadE006ret33, thus it mimics a fly-by encounter scenario. Although the actual value of the off-set between the stellar disc and the DM halo in this model is smaller, the off-set persists till the very end of the simulation run due to the continued pericenter passages of the satellite galaxy.
\par
Lastly, we probe the temporal evolution of the disc-DM halo off-set in the in-plane ($x-y$ plane) and in the vertical direction ($x-z$ plane) for different minor merger models. We found that the in-plane separation/off-set varies in the range $\sim 300 -500 \pc$ (see Fig.~\ref{fig:sloshing_of_centres}). However, the separation in the direction perpendicular to the disc mid-plane is less than the softening length ($200 \pc$) of the simulation, hence they are not shown here. This trend remains true for all minor merger models considered here.
\par
In appendix~\ref{sec:isolatedEvolution}, we show the temporal evolution of the disc-DM halo separation for the isolated gSa galaxy model (hereafter isogSa). The galaxy model, when evolved in isolation, does not produce an off-set between the centres of the disc and the DM halo. This accentuates the fact that a pericenter passage of a satellite galaxy can drive a transient off-centred disc-DM halo configuration for a wide variety of orbital configurations considered here. A similar scenario of generating a transient off-set between disc and DM halo via dwarf-dwarf merger has been shown in \citet{Pardyetal2016}.
\par
 For the sake of completeness, we also examine whether a similar off-set is generated between the stellar disc and the bulge of the host galaxy. This is also shown in Fig.~\ref{fig:fig2}. As seen clearly, the disc and the bulge of the host galaxy always remain concentric, and this holds true for different orbital configurations considered here.

\section{Quantifying lopsided asymmetry in the density distribution}
\label{sec:Lopsidedness}

\begin{figure*}
\centering
\textbf{gSadE001dir33}
\includegraphics[width=\linewidth]{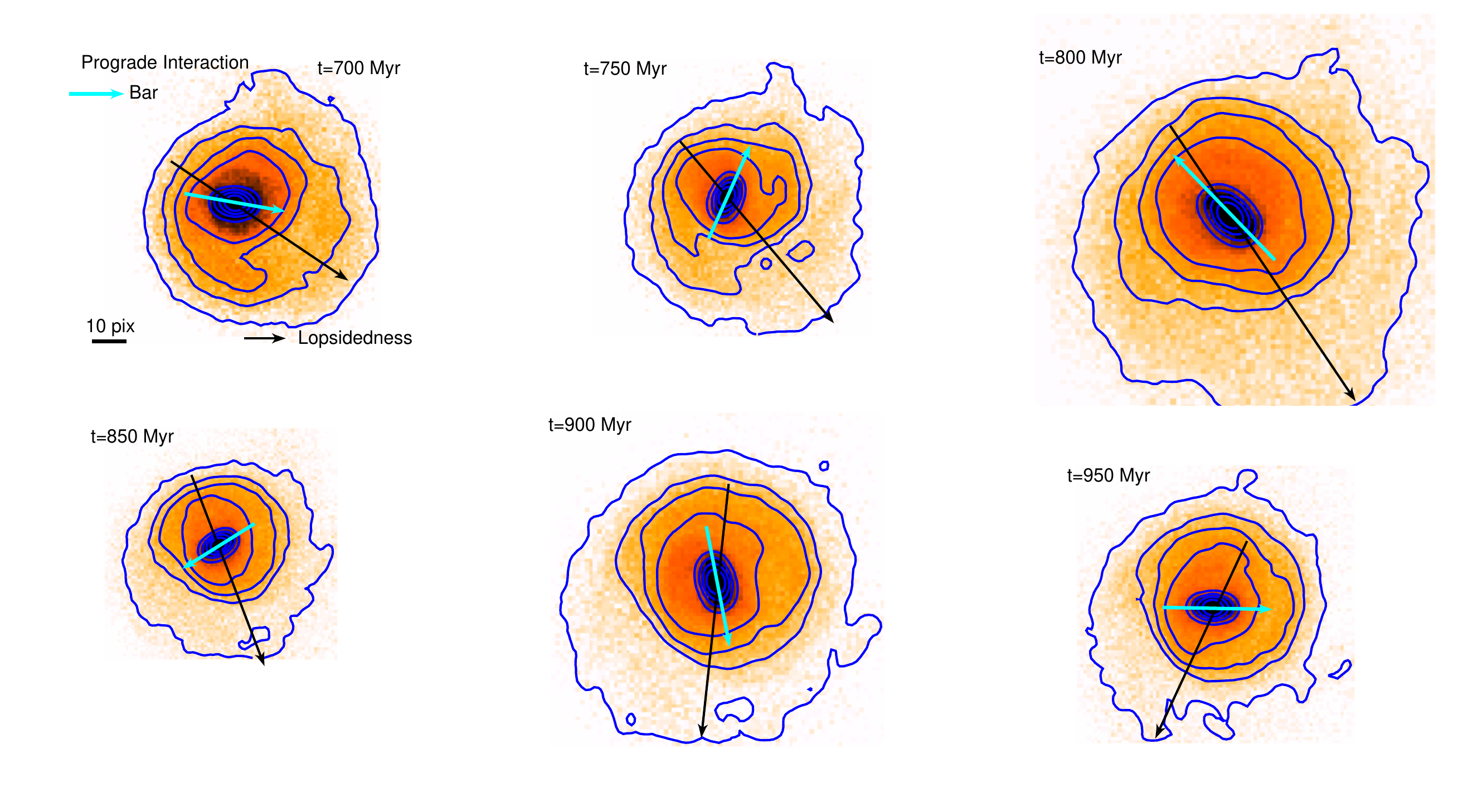}
\medskip
\vspace{0.5 cm}
\textbf{gSadE006ret33}
\includegraphics[width=\linewidth]{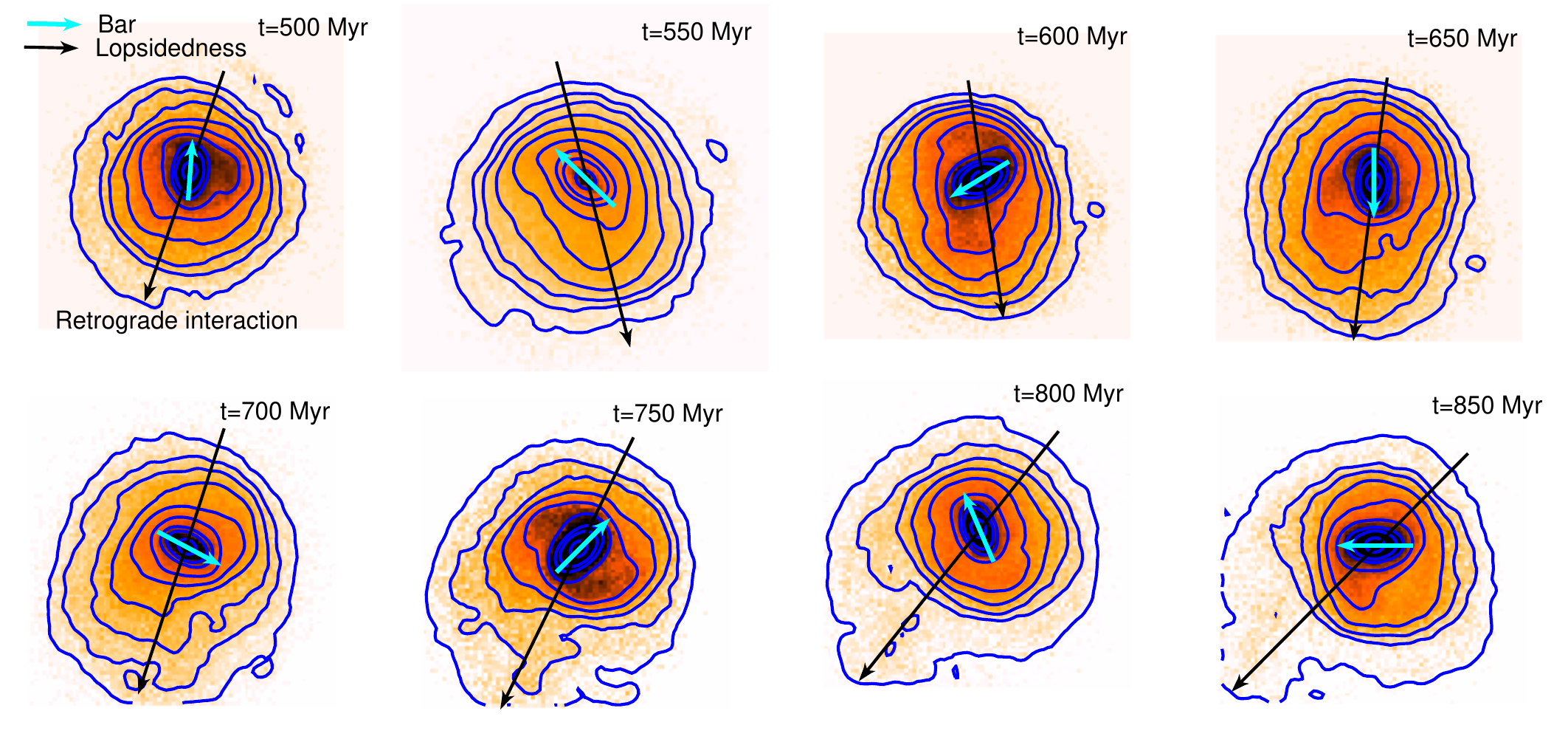}
\caption{ Face-on density distribution of the host galaxy's disc particles are shown for models gSadE001dir33 (top panels) and gSadE006ret33 (bottom panels), during one full rotation of the bar after the first pericenter passage happens. The position angles of the bar and the $m=1$ lopsidedness are denoted by arrows (of different colours, as shown at the top of each panel). Time between each snapshot is $50 \Myr$. Solid lines denote the contours of constant surface density.}
\label{fig:densmap_lopsided}
\end{figure*}

\begin{figure*}
\centering
\includegraphics[width=0.95\linewidth]{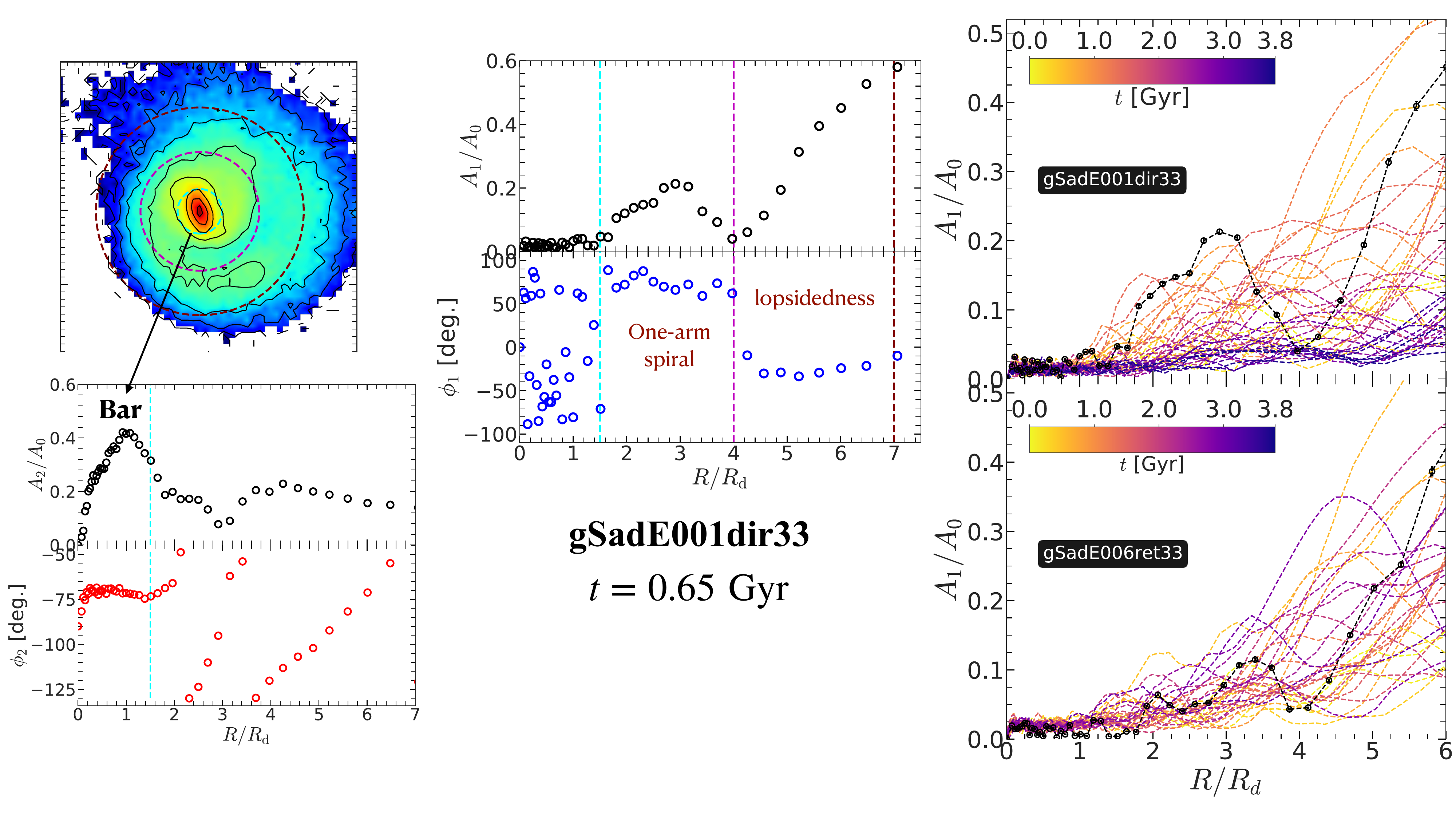}
\caption{Left panels show an example of coexisting one-arm spiral and an outer $m=1$ disc lopsidedness from the radial profiles of $A_1/A_0$ and the corresponding phase-angle ($\phi_1$). Similarly, the determination of bar extent from the amplitude and the phase-angle of the $m=2$ Fourier mode is also shown. The cyan, magenta, and the maroon circles denote the radial extents of $1.5 R_{\rm d}$, $4 R_{\rm d}$, and $7 R_{\rm d}$, respectively. Right panels show the radial profiles of $A_1/A_0$, calculated at different times, for two minor merger models gSadE001dir33 (top right panel) and gSadE006ret33 (bottom right panel). The colour bar denotes the epochs of the minor merger models.}
\label{fig:lopsided_radial}
\end{figure*}

Fig.~\ref{fig:density_illustration_collage} already indicated the existence of an $m=1$ asymmetry/distortion in the stellar density distribution of the host galaxy. This $m=1$ distortion is most prominent after each pericenter passage of the satellite. In this section, we study in detail the generation of the $m=1$ distortion in the stellar density distribution, identify the nature of the $m=1$ distortion as lopsidedness, and characterize its strength, and longevity.

Fig.~\ref{fig:densmap_lopsided} (top panels) shows the face-on density distribution of the host galaxy's disc particles for the minor merger model gSadE001dir33, at different time-steps after the first pericenter passage of the satellite galaxy. The existence of an $m=1$ asymmetry is clearly seen in the density maps. To quantify further, we calculate the radial variation of the $m=1$ Fourier harmonics of the stellar disc's density distribution of the host galaxy at different times. This is shown in Fig.~\ref{fig:lopsided_radial} (top right panel). 

\subsection {Characterizing the \textit{m=1} lopsided distortion in stars}
\label{sec:lopsided_distortions}
Here, we characterize the properties of an $m=1$ lopsidedness in the density distribution of the disc's stars of the host galaxy. This characterization is based on the amplitude ($A_1/A_0$) and the phase angle ($\phi_1$) obtained from the Fourier decomposition of the stellar density distribution. First we note that, a tail-like feature is produced due to the tidal pull during and (shortly) after a pericenter passage (e.g. see in the top panels of Fig~\ref{fig:densmap_lopsided}). This tidal tail, in turn, yields a large non-zero value of the $m=1$ Fourier coefficient in the outer radial extent ($R > 7 R_{\rm d}$). Hence, this radial extent should be avoided while characterizing the $m=1$ lopsidedness in the density distribution.
\par
Next, we find that in our chosen model gSadE001dir33, a one-arm spiral forms after the first pericenter passage of the satellite (e.g. see at $t = 0.7-0.75 \Gyr$ in Fig~\ref{fig:densmap_lopsided}). The same one-arm spiral reappears after the second pericenter passage of the satellite as well. Eventually it fades away after $\sim 200 \Myr$. We check that the presence of such a one-arm spiral produces a hump-like feature in the radial profile of the $m=1$ Fourier coefficient (see the example shown in the left panels of Fig.~\ref{fig:lopsided_radial}). The extent of this one-arm spiral varies from $\sim 2-4 R_{\rm d}$ in the model gSadE001dir33 \footnote{Although, sometimes it can extend beyond the radial extent mentioned here.}. Further, Fig.~\ref{fig:lopsided_radial} (middle panel) revealed that, within the radial extent $4-7 \ R{\rm d}$, the Fourier coefficient of the $m=1$ Fourier harmonics ($A_1/A_0$) increases with radius; thereby indicating the presence of an outer $m=1$ disc lopsidedness in the model gSadE001dir33. The corresponding radial variation of the phase angle ($\phi_1$), calculated at $t =0.65 \Gyr$, is also shown in Fig.~\ref{fig:lopsided_radial} (see middle panel). We notice that the phase-angle ($\phi_1$) values for the one-arm spiral dominated region and the outer $m=1$ disc lopsidedness are very different, thus indicating presence of two lopsided patterns with different phase angles.
\par
The radial variation of the phase-angle associated to an $m=1$ lopsidedness is important for characterizing its nature and determining its physical origin \citep[for details see][]{JogandCombes2009}. We calculated the mean and the corresponding standard deviation (SD hereafter) of the  phase angle ($\phi_1$), at $t =0.65 \Gyr$, for these two above-mentioned regions. We find that, in the region hosting a prominent one-arm spiral ($\sim 2-4 R_{\rm d}$), the mean value of $\phi_1$ is $\sim 72 \degrees$ with an associated SD of $\sim 9 \degrees$. On the other hand,  in the region of outer $m=1$ disc lopsidedness, the mean value of $\phi_1$ is $\sim -29  \degrees$ with an associated SD of $\sim 6.8 \degrees$. We further note that in a self-gravitating disc, the survival of an $m=1$ lopsidedness depends strongly on the differential precession $\Omega - \kappa$ \citep[for details see, e.g.,][]{Sahaetal2007,JogandCombes2009} which is, in general, non-zero for a self-gravitating disc (unlike a Keplerian disc where $\Omega = \kappa$). Here, $\Omega$ and $\kappa$ are the circular and the epicyclic frequencies, respectively. \citet{Baldwinetal1980} estimated the time-scale of winding up for an $m=1$ lopsidedness as $\tau_{\rm lop} = 2 \pi/ \Delta (\kappa - \Omega)$. We checked that, for the one-arm spiral region, $\kappa - \Omega \sim 18 \kms$ kpc$^{-1}$, which in turn, gives $\tau_{\rm lop} \sim 350$ Myr, so the pattern would wind up rather quickly. On the other hand, for the region displaying outer $m=1$ disc lopsidedness, the $\kappa - \Omega$ value is close to zero, thereby producing a large $\tau_{\rm lop}$. In other words, the outer $m=1$ disc lopsidedness will not wind up so quickly when compared to the one-arm spiral. So, from this point on, we will only focus on the properties, temporal variation and the pattern rotation of the outer $m=1$ disc lopsidedness ($\sim 4-7 R_{\rm d}$).
\par
As mentioned before, the outer $m=1$ lopsided distortion exists predominantly within $\sim 4-7 R_{\rm d}$ \footnote {It should be noted that the choice of $4-7 R_{\rm d}$ to quantify the strength of the $m=1$ lopsidedness is not generic, the extent of the $m=1$ lopsidedness varies from galaxy to galaxy, and depends on the properties of a galaxy.} at different times in the model gSadE001dir33. Also, for the outer $m=1$ disc lopsidedness, the corresponding strength of the $m=1$ lopsidedness increases with radius (as denoted by the increasingly higher values of $A_1/A_0$ with radius, see e.g., in Fig.~\ref{fig:lopsided_radial}). This trend is in agreement with the observational studies of the $m=1$ lopsidedness in galaxies \citep[e.g., see][]{RixandZaritsky1995,RudnickandRix1998,Conseliceetal2000,Angirasetal2006,Reichardetal2008,vanEymerenetal2011,Zaritskyetal2013}. We also checked that, after the merger of the satellite, the stellar particles from the satellite do not contribute appreciably to a coherent $m=1$ lopsidedness in the density distribution; hence, they are discarded in the subsequent analyses.

\subsection {Temporal variation of the \textit{m=1} lopsided distortion}
\label{sec:temp_lopsided_distortions}
 Next, we study the temporal evolution of the $m=1$ lopsidedness in the stellar density distribution of the host galaxy for the minor merger models considered here. For the model gSadE001dir33, initially ($t=0$) there was no discernible lopsidedness in the outer part as inferred from the $A_1/A_0$ value close to zero in the outer disc regions (see top right panel of Fig.~\ref{fig:lopsided_radial}). A prominent coherent lopsidedness appears only after the first pericenter passage of the satellite; the average $A_1/A_0$ value reaches close to $0.4$; followed by a decrease in the $A_1/A_0$ value as the satellite moves farther away after the pericenter passage. The strong lopsidedness reappears after the second pericenter passage of the satellite galaxy, marked by an increase in the average $A_1/A_0$ value in the outer disc region ($\sim 4.5 -6 R_{\rm d}$). However, after the satellite merges with the host galaxy, the lopsidedness gets weakened subsequently in the post-merger remnant. By the end of the simulation run, at $t = 3.8 \Gyr$, the $A_1/A_0$ value becomes less than 0.1 in the outer parts ($\sim 4-7 R_{\rm d}$) of the stellar disc (see top right panel of Fig.~\ref{fig:lopsided_radial}), thereby denoting the absence of a strong, coherent lopsidedness. 
 \par
\begin{figure}
\centering
\includegraphics[width=1.0\linewidth]{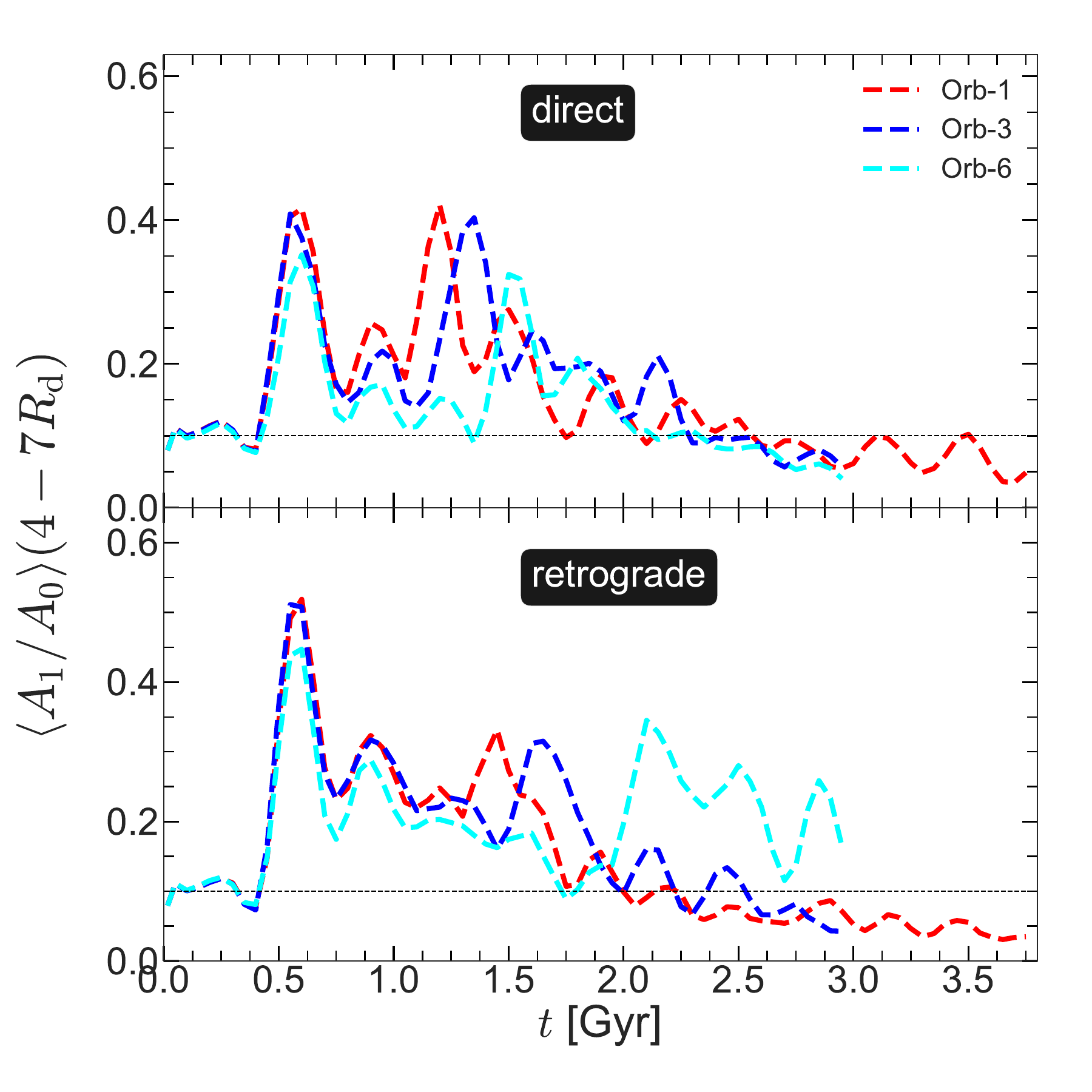}
\caption{Evolution of the median value of the $m=1$ Fourier coefficient $A_1/A_0$, calculated within the radial extent of $4-7 R_{\rm d}$, is shown as a function of time for several minor merger models with direct (top panel) and retrograde (bottom panel) configurations. The black horizontal line denotes $\avg{A_1/A_0}=0.1$, and is used as a demarcation for the onset of the $m=1$ lopsidedness \citep{JogandCombes2009}.}
\label{fig:lopsised_temporalEvolution}
\end{figure}

 In order to make a systematic study on the temporal evolution of the $m=1$ lopsidedness for all models, next we study how the values of $A_1/A_0$ vary with time. To do that, for a certain model and at time $t$, we first compute the median value of $A_1/A_0$, within the radial extent $4-7 R_{\rm d}$ where the lopsidedness is prominent, and then study the resulting temporal variation.  
Fig.~\ref{fig:lopsised_temporalEvolution} shows the corresponding temporal variation of the $A_1/A_0$ values for six merger models with different orbital configurations (direct and retrograde) \footnote {We have computed the temporal evolutions for all minor merger models considered here. For the clarity of the representation, we are showing for only 6 models here.} As seen clearly from Fig.~\ref{fig:lopsised_temporalEvolution}, the broad trend of the temporal variation of the $m=1$ lopsided distortion, for all 5 minor merger models shown here (except gSadE006ret33), is similar to what is shown for the model gSadE001dir33.
\par
 The question remains what happens to the sustainability of the $m=1$ lopsided feature when the merger happens at a very late epoch and the host galaxy experiences continued pericenter passage of the satellite? To investigate that, we select the minor merger model gSadE006ret33 where the merger happens at a later epoch (see Table.~\ref{table:key_param}). Fig.~\ref{fig:densmap_lopsided} (bottom panels) shows the face-on density distribution of the host galaxy's disc particles for the model gSadE006ret33, at different time-steps after the first pericenter passage of the satellite galaxy. Also, the corresponding radial variations of the $m=1$ Fourier harmonics of the density distribution are shown in Fig.~\ref{fig:lopsided_radial} (bottom right panel).  Fig~\ref{fig:lopsised_temporalEvolution} (bottom panel) demonstrates that as the host galaxy experiences continued pericenter passages, the lopsided pattern also persists till the end of the simulation run ($t = 3 \Gyr$); the average values of $A_1/A_0$ in the outer disc region ($\sim 4 -7 R_{\rm d}$) are non-zero, and are higher than values calculated at $t=0$. This demonstrates that, although the $m=1$ lopsided distortion excited by a single pericenter passage does not last long, the continued pericenter passages can maintain the net lopsided feature in the host galaxy for a longer time-scale.
 \par
 In appendix~\ref{sec:isolatedEvolution}, we show that when the host galaxy model is evolved in isolation, no prominent $m=1$ lopsided distortion gets excited. The average values of $A_1/A_0$ throughout the entire disc region remain below 0.1, indicating the absence of a strong $m=1$ lopsidedness in the stellar density distribution of the host galaxy. This reinforces the fact that minor merger event is liable to trigger the lopsided disturbance in the host galaxy.
 \begin{figure}
\centering
\includegraphics[width=1.0\linewidth]{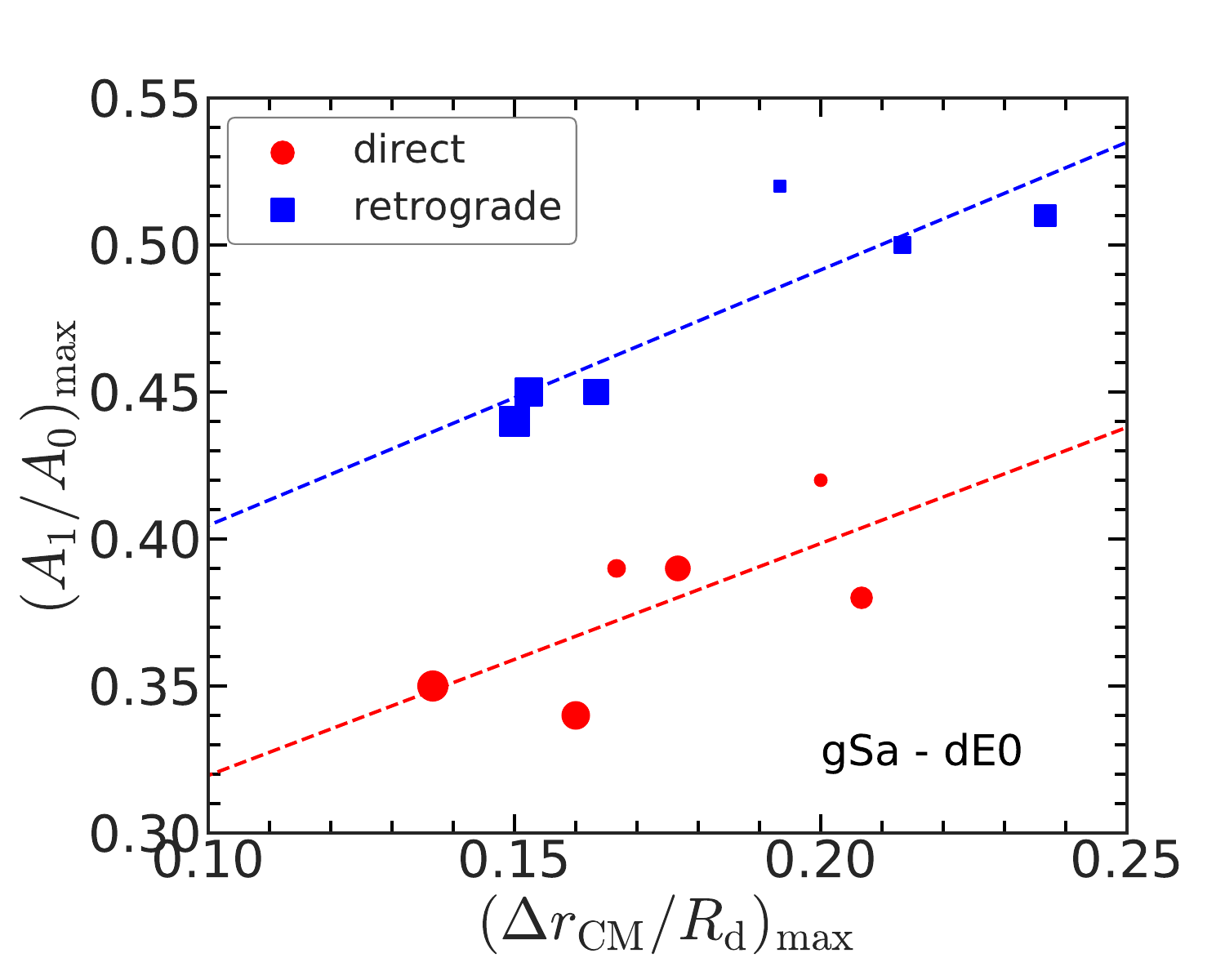}
\caption{Maximum values of the $m=1$ Fourier coefficient ($A_1/A_0$) and the separation between the centres of stellar disc and the DM halo ($\Delta r_{\rm CM}$) are shown for different minor mergers (of gSa-dE0 type) with direct (red circles) and retrograde (blue squares) configurations. Dashed lines denote the corresponding best-fit straight lines to the points. For a direct orbital configuration, the values are calculated just after the first pericenter passage while for a retrograde orbital configuration, the values are calculated just after the second pericenter passage (for details, see text). The increasing size of the points denote higher orbit numbers, for details see section~\ref{sec:simu_setup}. Here, $R_{\rm d} = 3 \kpc$.}
\label{fig:lopsided_comparison_offset}
\end{figure}
\par
 Also, we notice a correlation in the formation epoch of a stellar disc-DM halo off-set and the excitation of the lopsided pattern in the host galaxy (compare bottom panels of Fig.~\ref{fig:density_illustration_collage} and Fig.~\ref{fig:densmap_lopsided}). Here, we compare the maximum values of the $m=1$ Fourier coefficient and separation between the centres of stellar disc and the DM halo ($\Delta r_{\rm CM}$) for different minor merger models with direct and retrograde configurations. We checked that for the minor merger models in the direct orbital configuration, the maxima for both the $\Delta r_{\rm CM}$ and the $\left<A_1/A_0\right >$ occur after the first pericenter passage. Similarly, for models in the retrograde orbital configuration, the maxima for both the $\Delta r_{\rm CM}$ and the $\left<A_1/A_0\right >$ occur after the second pericenter passage. We then compare these maximum values, computed at the above-mentioned epochs, for the models considered here. This is shown in Fig.~\ref{fig:lopsided_comparison_offset}. As seen clearly, for both orbital spin configurations, these two maximum values follow a (nearly) linear relation. This is not surprising since both physical phenomena are driven by the tidal forces exerted on the host galaxy by the satellite during the pericenter passages.

\section {Pattern speed measurement and resonances}
\label{sec:pattern_speed}

\begin{figure*}
\centering
\includegraphics[width=0.85\linewidth]{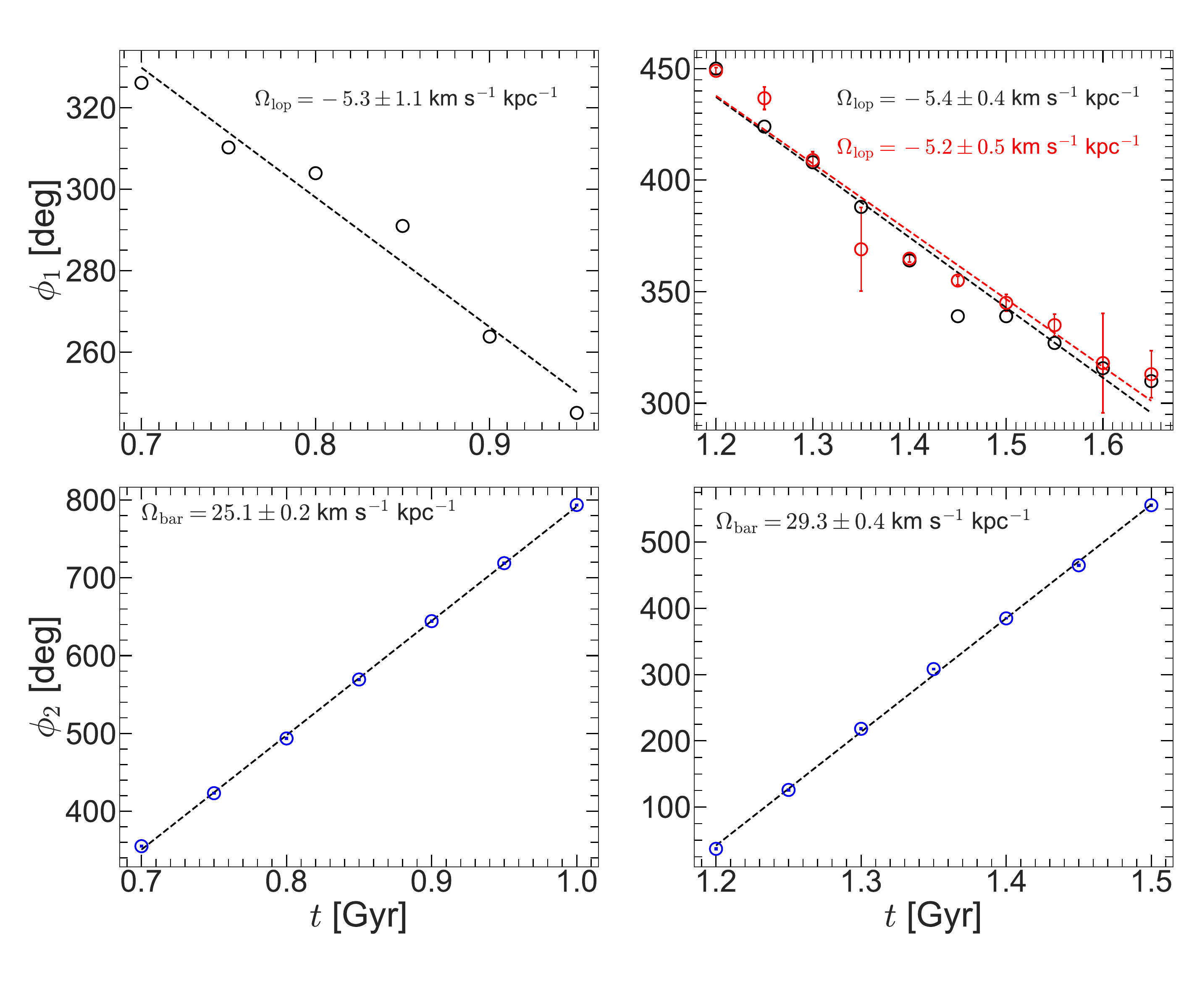}
\caption{Pattern speed measurement of the $m=1$ lopsided distortion (top panels) and the $m=2$ bar mode (bottom panels) are shown at two epochs after the first and second pericenter passages for the model gSadE001dir33. Black dashed lines denote the best-fit straight line to the temporal variation of the phase-angle (for the bar) as well as the temporal variation of the orientation of the density isophotal contours (for the lopsidedness). Red points and the red dashed line in the top right panel denote the measurement of the pattern speed of the $m=1$ lopsidedness via fitting a straight line to the temporal variation of the phase-angle ($\phi_1$). Measured pattern speed values are indicated in each sub-panel.}
\label{fig:lopsided_patternspeed}
\end{figure*}

A visual inspection of Fig.~\ref{fig:densmap_lopsided} already provided the indication that the $m=1$ lopsided pattern rotates in the disc, in a similar fashion the $m=2$ bar mode rotates. Here, we measure simultaneously the pattern speeds of the $m=1$ lopsided distortion as well as the central $m=2$ bar mode in the model gSadE001dir33. For that, we choose two time-intervals of $\sim 0.3 \Gyr$, after the first and the second pericenter passages of the satellite when both the $m=1$ lopsided distortion and the $m=2$ bar mode co-exist. This simultaneous measurements will ease the determination of the direction of the pattern speed of the $m=1$ lopsidedness with respect to the $m=2$ bar pattern speed. The bar pattern speed ($\Omega_{\rm bar}$) is measured by fitting a straight line to the temporal variation of the phase-angle ($\phi_2$) of the $m=2$ Fourier mode. This assumes that the bar rotates rigidly with a single pattern speed in that time-interval. The resulting measurements of the $m=2$ bar pattern speeds ($\Omega_{\rm bar}$) are shown in Fig.~\ref{fig:lopsided_patternspeed} (bottom panels). The pattern speed of the $m=1$ lopsided distortion ($\Omega_{\rm lop}$) is measured by fitting a straight line to the  time variation of the orientation of the density isophotal contours. This is shown in Fig.~\ref{fig:lopsided_patternspeed} (top panels). We note that, when the $m=1$ lopsided distortion rotates with a well-defined, single pattern speed (similar to an $m=2$ bar mode), it is possible to measure the pattern speed of the $m=1$ lopsided distortion by fitting a straight line to the temporal variation of the phase-angle ($\phi_1$) of the $m=1$ lopsidedness. However, we show that these two measurements of the pattern speed of the lopsidedness match pretty well within their error limits (see top right panel of Fig.~\ref{fig:lopsided_patternspeed}).
\par
The simultaneous measurements of the bar and the lopsided pattern speeds reveal two important aspects. First, the bar rotates faster than the $m=1$ lopsided distortion. For example, around $t = 1.4 \Gyr$, the bar pattern speed ($\Omega_{\rm bar}$) is $29.3 \pm 0.4 \kms$ kpc$^{-1}$ whereas around the same epoch, the $m=1$ lopsided distortion rotates with a pattern speed ($\Omega_{\rm lop}$) of $-5.2 \pm 0.5 \kms$ kpc$^{-1}$. This trend also holds for the chosen time-interval after the first pericenter passage of the satellite (compare left panels of Fig.~\ref{fig:lopsided_patternspeed}).  Secondly, the $m=1$ lopsided distortion is in retrograde motion with respect to the bar rotation as well as the underlying disc rotation. 
\par
\begin{figure}
\centering
\includegraphics[width=1.0\linewidth]{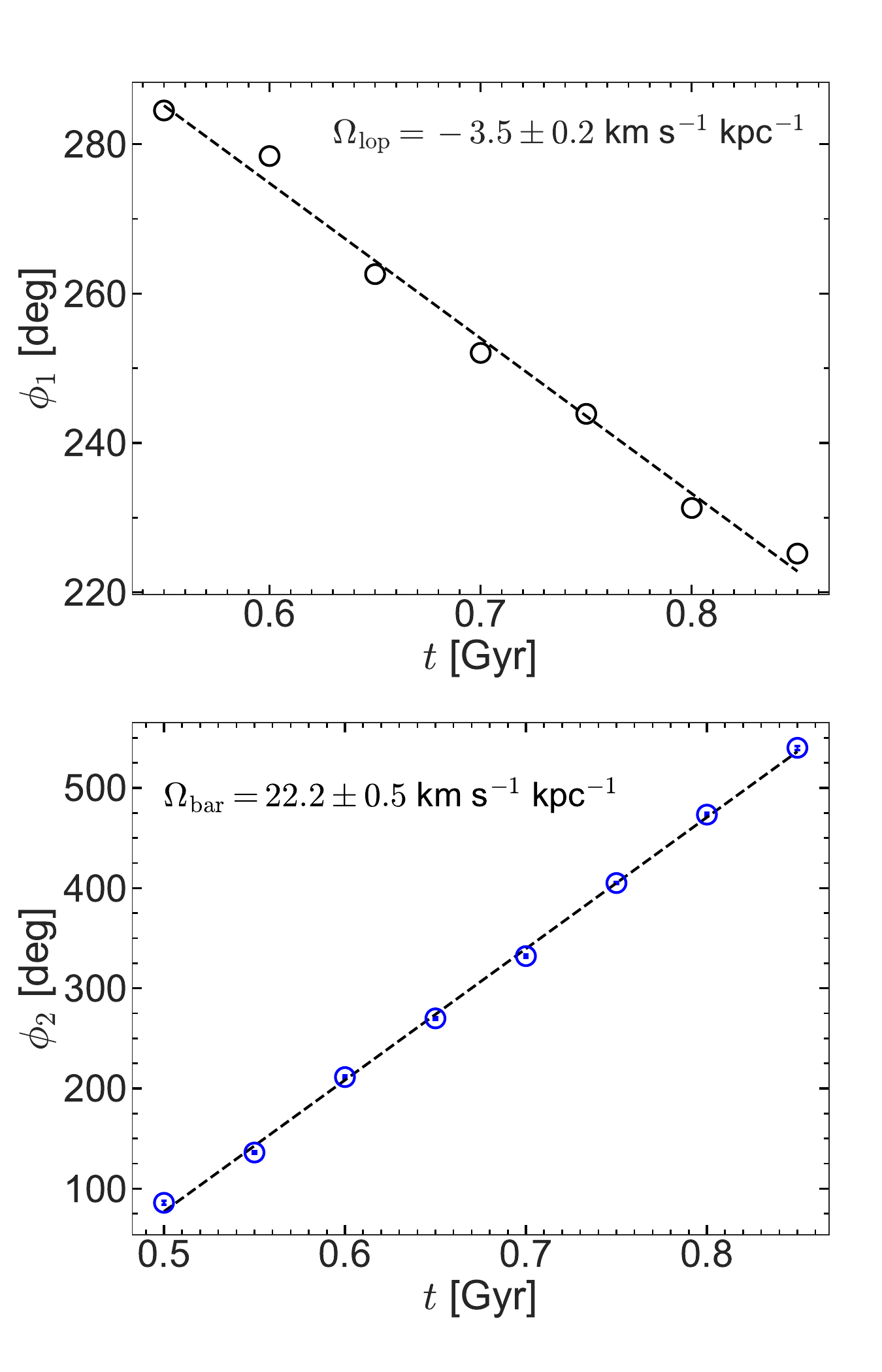}
\caption{Pattern speed measurement of the $m=1$ lopsided distortion (top panel) and the $m=2$ bar mode (bottom panel) are shown after the first pericenter passage of the satellite for the model gSadE006ret33. Measured pattern speed values are indicated in each sub-panel.}
\label{fig:lopsided_patternspeed_ret}
\end{figure}

Lastly, we investigate whether the direction of the orbital spin vector plays any role in deciding the sense of the rotation of the $m=1$ lopsidedness with respect to the $m=2$ bar mode. To achieve that, we choose the model gSadE006ret33. In Fig.~\ref{fig:densmap_lopsided} (bottom panels), the density distributions of the stellar particles after the first pericenter already indicated a pattern rotation of the $m=1$ lopsided pattern, similar to what is seen in the model gSadE001dir33. Using the same methodology, as mentioned above, we simultaneously measure the pattern speeds of the $m=2$ bar mode and the $m=1$ lopsided pattern after the first pericenter passage of the satellite. This is shown in Fig.~\ref{fig:lopsided_patternspeed_ret}. For the retrograde orbital configuration also, the lopsided pattern rotates much slower ($\Omega_{\rm lop} =-3.5 \pm 0.2 \kms$ kpc$^{-1}$  at $t \sim 0.75 \Gyr$) than the bar ($\Omega_{\rm bar} =22.2 \pm 0.5 \kms$ kpc$^{-1}$  at $t \sim 0.75 \Gyr$), and the sense of rotation is in retrograde with respect to the bar. The physical implications of the retrograde pattern speed (with respect to the bar) of the lopsidedness in our minor merger models is discussed below.

\par
\begin{figure}
\centering
\includegraphics[width=1.\linewidth]{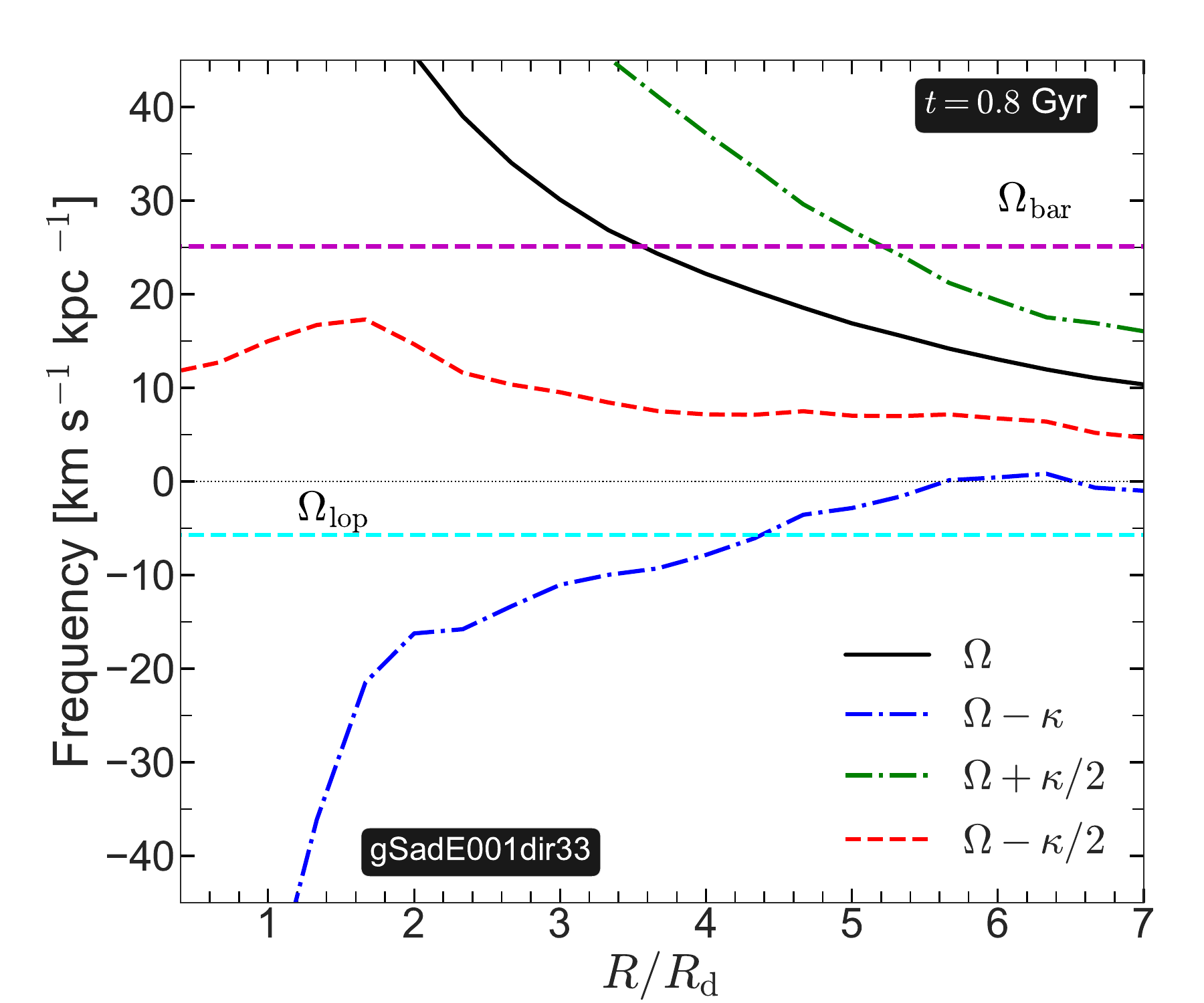}
\caption{ The circular and the epicyclic frequencies, together with the locations of the resonant points of the $m=2$ bar and $m=1$ lopsideness are shown at $t = 0.8 \Gyr$ for the model gSadE001dir33. The magenta dashed horizontal line denotes the pattern speed ($\Omega_{\rm bar}$) of the bar while the cyan dashed horizontal line denotes the pattern speed ($\Omega_{\rm lop}$) of the $m=1$ lopsidedness.}
\label{fig:frequencies}
\end{figure}

Following \citet{BiineyTremaine2008}, the dispersion relation for a tightly wound $m=1$ pattern in a disc  can be written as
\begin{equation}
    (\omega - \Omega)^2 = \kappa^2(R) -2\pi G \Sigma_{0}(R) |k| + \sigma^2 k^2,
\end{equation}
\noindent where $\Sigma_{0}$ is the surface density of the disc, $\Omega$ and $\kappa$ are the angular and radial epicyclic frequencies; $\sigma$ refers to the velocity dispersion and $k$ is the wavenumber. In the absence of self-gravity, the relevant free precession frequency corresponding to the $m=1$ mode in a cold disc is $\omega = \Omega - \kappa$ at given radius $R$. Since in a realistic galaxy model with stellar disc and dark matter halo, $\kappa > \Omega$, and hence $\omega = \Omega - \kappa < 0$ at all radii \cite[see][for various mass models]{SahaandJog2014}. As shown in Fig.~\ref{fig:frequencies}, the values of $\Omega - \kappa$ remain less than 0, for almost all radii considered here, at $t = 0.8 \Gyr$. At this point, we caution the reader that, the $m=1$ lopsidedness is shown to have roughly constant phase angle over a range of radii, and hence, the application of the WKB or the tight-winding approximation is not rigorously valid. However, even when $|kR| \sim 1$ or $0.5$, the WKB or the tight-winding approximation provides a valuable insight on the nature of the  perturbation, especially to predict a rough guideline for the expected frequency of the perturbation \citep[for details, see][]{BiineyTremaine2008}. In the same spirit, we applied WKB approximation here to get a broad understanding for the frequencies of different modes.
 It is worth mentioning that in previous studies, \citet{JunqueiraCombes1996,Baconetal2001} have shown excitation of an $m=1$ lopsidedness in the central region of galaxies e.g., M~31 nucleus and the corresponding pattern speeds are positive, generally high, since dynamical time is also shorter inside. These central $m=1$ modes mimic well the pressure mode ($p$-mode) as described in the context of near-Keplerian disc by \cite{Tremaine2001}. However, the $m=1$ lopsidedness that we measure in our current simulation set-up are dominated by self-gravity (more like the $g$-modes) and hence the pattern speed is expected to be following the $\Omega-\kappa$ curve in the galaxy model chosen. In the outer part ($\sim 4 - 7 R_d$) of our galaxy models where we measure lopsidedness, the absolute value of $\Omega -\kappa$ is smaller and we also obtain comparatively smaller pattern speed. Interestingly, in the case of a pure exponential stellar disk, the value of $\Omega-\kappa \sim 0$ beyond about $5R_d$ and it switches sign from negative to positive at $4.6R_d$ \citep{SahaandJog2014}. In pure disk only simulations, it was shown that the outer lopsidedness had pattern speed close to zero as well \citep{Sahaetal2007}.
\par
In our current measurements (see Fig.~\ref{fig:lopsided_patternspeed}), we show that lopsidedness in the outer parts of our galaxy model has negative pattern speed, with $\Omega_{\rm lop} = -5.4 \kms$ kpc$^{-1}$ after the second pericenter passage in model gSadE001dir33. The pattern speed is negative after the first pericenter passage as well. For the retrograde model gSadE006ret33 (Fig.~\ref{fig:lopsided_patternspeed_ret}), we measure $\Omega_{\rm lop} = -3.5 \kms$ kpc$^{-1}$ after the first pericenter passage. Since the pattern speed is negative, our galaxy models do not have a cororation resonance for the lopsidedness but only the inner Lindblad resonance (ILR). For the model gSadE001dir33, $\Omega_{\rm lop} = -5.3 \kms$ kpc$^{-1}$, the ILR for the lopsidedness is pushed to $4.3 R_{\rm d}$, towards the outer region of the stellar disc (see Fig.~\ref{fig:frequencies}). In the same galaxy model, the central bar has a positive pattern speed and the corotation for the bar in model gSadE001dir33 is well inside the disc, at $3.5 R_{\rm d}$. It is tantalising to infer that the central bar and the outer lopsidedness in our galaxy models are two dynamical patterns having different originating mechanisms. After all, the lopsidedness has been generated during the minor merger process while the central bar was present from the start of the minor merger simulation.
 However, the  ILR of the $m=1$ lopsidedness ($\sim 4.3 R_{\rm d}$) falls in between the CR ($\sim 4.2 R_{\rm d}$) and the OLR ($\sim 5.2 R_{\rm d}$) of the $m=2$ bar,  i.e., in the outer part of the stellar disc these resonance points almost overlap. Past studies have addressed the important role of resonance overlap due to bar-spiral or spiral-spiral scenarios \citep[e.g., see][]{SellwoodandBinney2002,MinchevandFamaey2010,Minchevetal2011} in the context of radial migration and disc dynamics. It will be insightful to investigate the stellar and gas kinematics at these (nearly overlapping) resonance locations associated with the $m=1$ lopsidedness and the $m=2$ bar, provided the $m=1$ lopsidedness is long-lived.

\section{Excitation of kinematic lopsidedness}
\label{sec:flapping_mode}

Past studies, both theoretically \citep[e.g., see][]{Jog1997,Jog2002} and from the observations \citep[e.g., see][]{vanEymerenetal2011,vanEymeren2011}, have shown that an $m=1$ lopsided distortion in the density distribution is associated with a large-scale asymmetry in the velocity field as well. An off-set between the rotation curves, calculated for the receding and approaching sides separately for a galaxy, is considered as the signature of a kinematic lopsidedness. In the light of these past findings, we now investigate the details of the kinematic lopsidedness in our selected models of minor merger.
\par
\begin{figure*}
\centering
\includegraphics[width=0.8\linewidth]{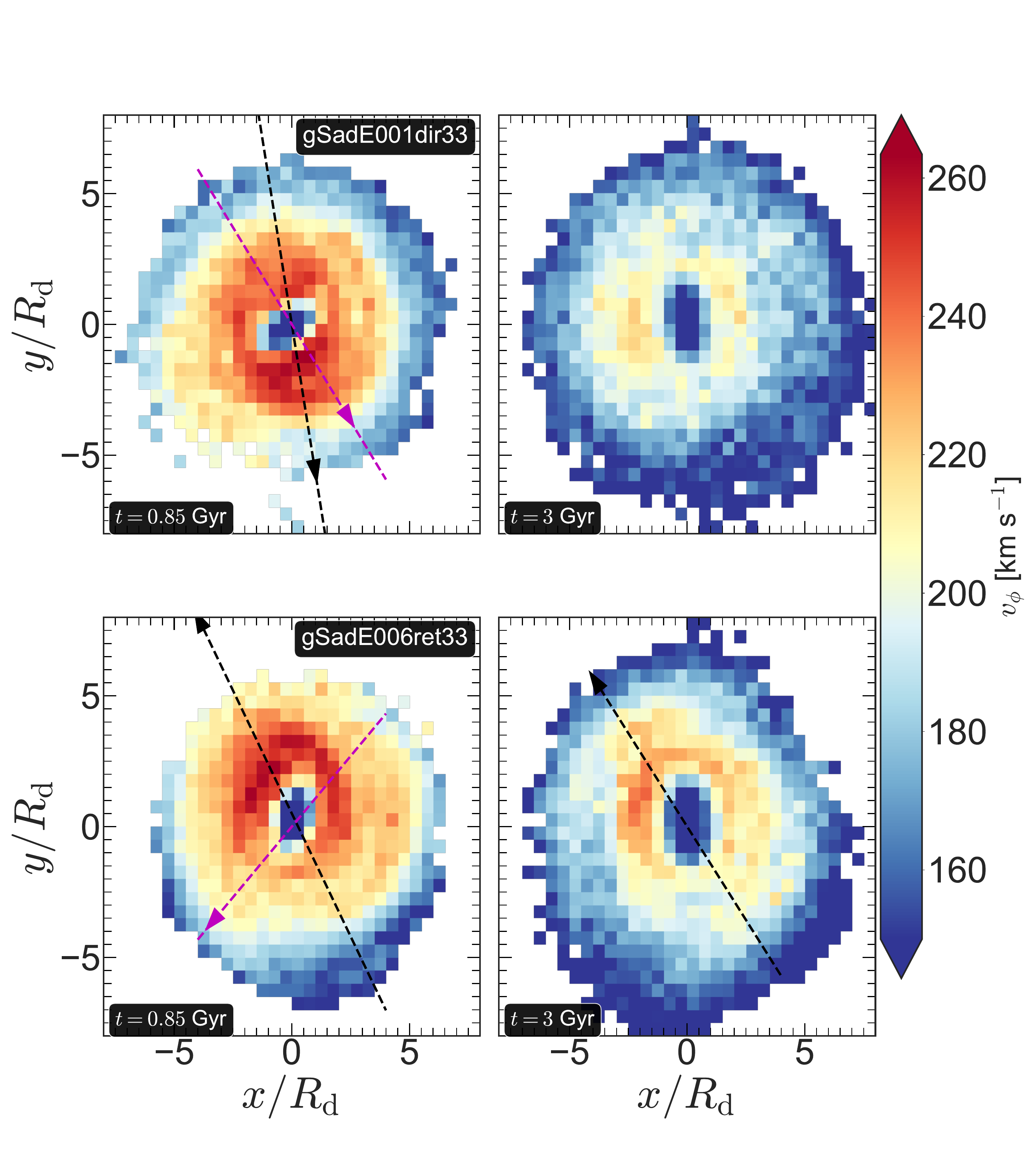}
\caption{ Top panels : distribution of the azimuthal velocity ($v_{\phi}$) in the $x-y$ plane, calculated at two different epochs for the model gSadE001dir33. Bottom panels show the corresponding distribution of the azimuthal velocity ($v_{\phi}$) in the $x-y$ plane, calculated at two different epochs for the model gSadE006ret33. The black arrow-line (when present) indicates the direction along which kinematic lopsidedness is prominent. The magenta arrow-line (when present) indicates the corresponding direction of the $m=1$ density lopsidedness.}
\label{fig:lopsided_kinematic}
\end{figure*}

To achieve that, we first choose the model gSadE001dir33. After the first pericenter passage of the satellite galaxy, this model displays a prominent large-scale $m=1$ lopsided distortion in the density distribution of the stellar disc in the host galaxy (see previous sections).  Using the intrinsic position-velocity information of the disc particles of the host galaxy, we first compute the distribution of the azimuthal velocity ($v_{\phi}$) in the $x-y$ plane. One such velocity distribution, calculated at $t = 0.85 \Gyr$, is shown in Fig.~\ref{fig:lopsided_kinematic} (see top left panel). A prominent, global distortion/asymmetry of the velocity distribution in the $x-y$ plane,  is seen to be present. To elaborate, if one extracts an one-dimensional velocity profile from this two-dimensional map, along the direction shown by an arrowed straight line (in black) in Fig.~\ref{fig:lopsided_kinematic}, the resulting one-dimensional velocity profile would show an asymmetry between positive and negative parts of $x$ values. This asymmetry is the signature of an $m=1$ kinematic lopsidedness. Also, we find a one-to-one correspondence between the epochs of prominent lopsidedness in the stellar density and the kinematics. This is not surprising since the kinematic lopsidedness can be thought as the representation of the $m=1$ density perturbation, but seen in velocity space. We further compute the distribution of the azimuthal velocity ($v_{\phi}$) in the $x-y$ plane at a later stage (say $t = 3 \Gyr$), when the satellite galaxy has merged with the host galaxy, and the post-merger remnant gets time to readjust itself. This is also shown in Fig.~\ref{fig:lopsided_kinematic} (top right panel). As seen clearly, the strong kinematic asymmetry that was present earlier at $t = 0.85 \Gyr$, is now disappeared, indicating that the prominent $m=1$ kinematic lopsidedness no longer persists. We check that the other minor merger models considered here, also show a similar trend of the temporal evolution of the kinematic lopsidedness in the host galaxy. For the sake of brevity, they are not shown here.

\begin{figure}
\centering
\includegraphics[width=\linewidth]{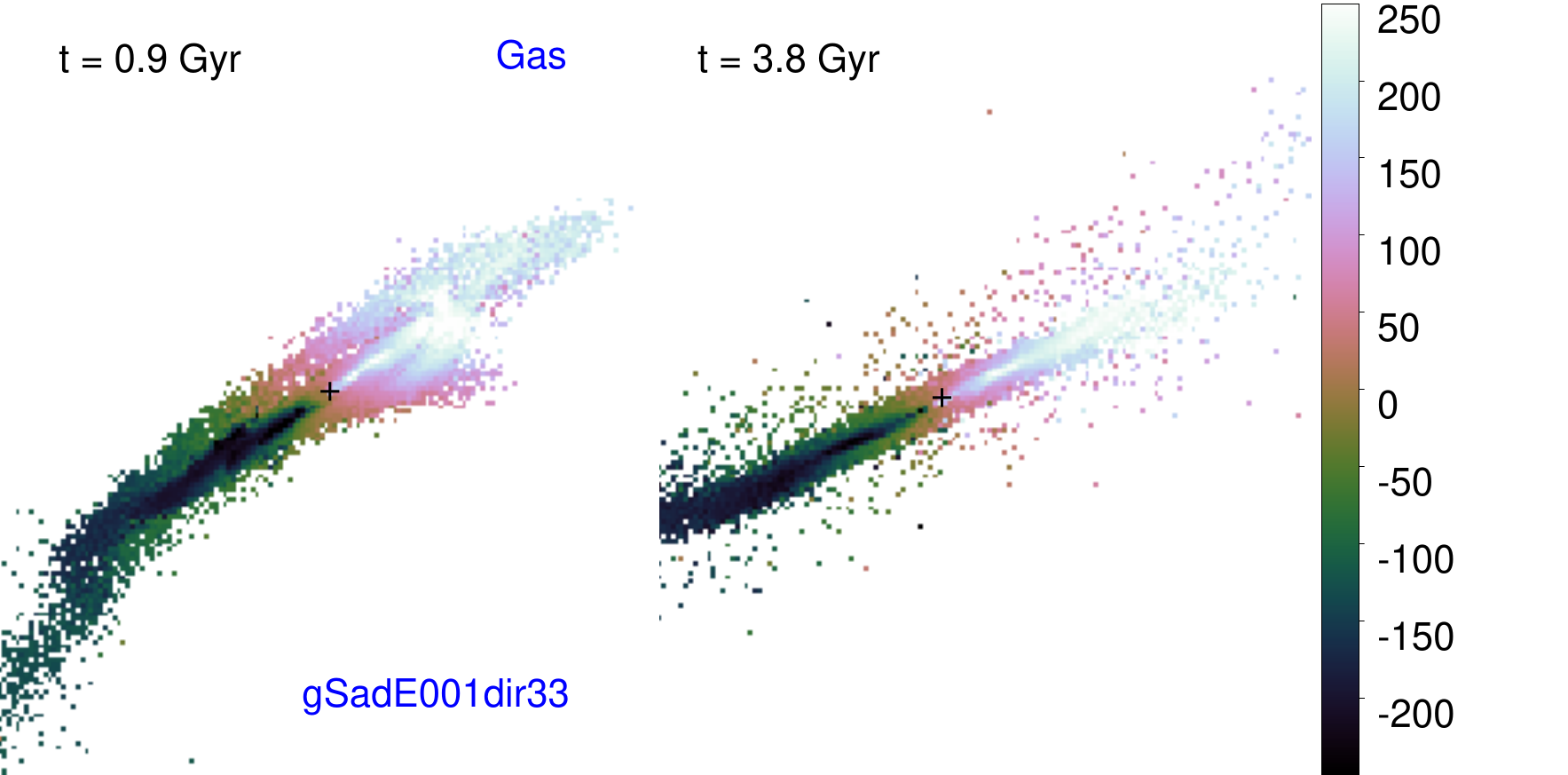}
\medskip
\includegraphics[width=\linewidth]{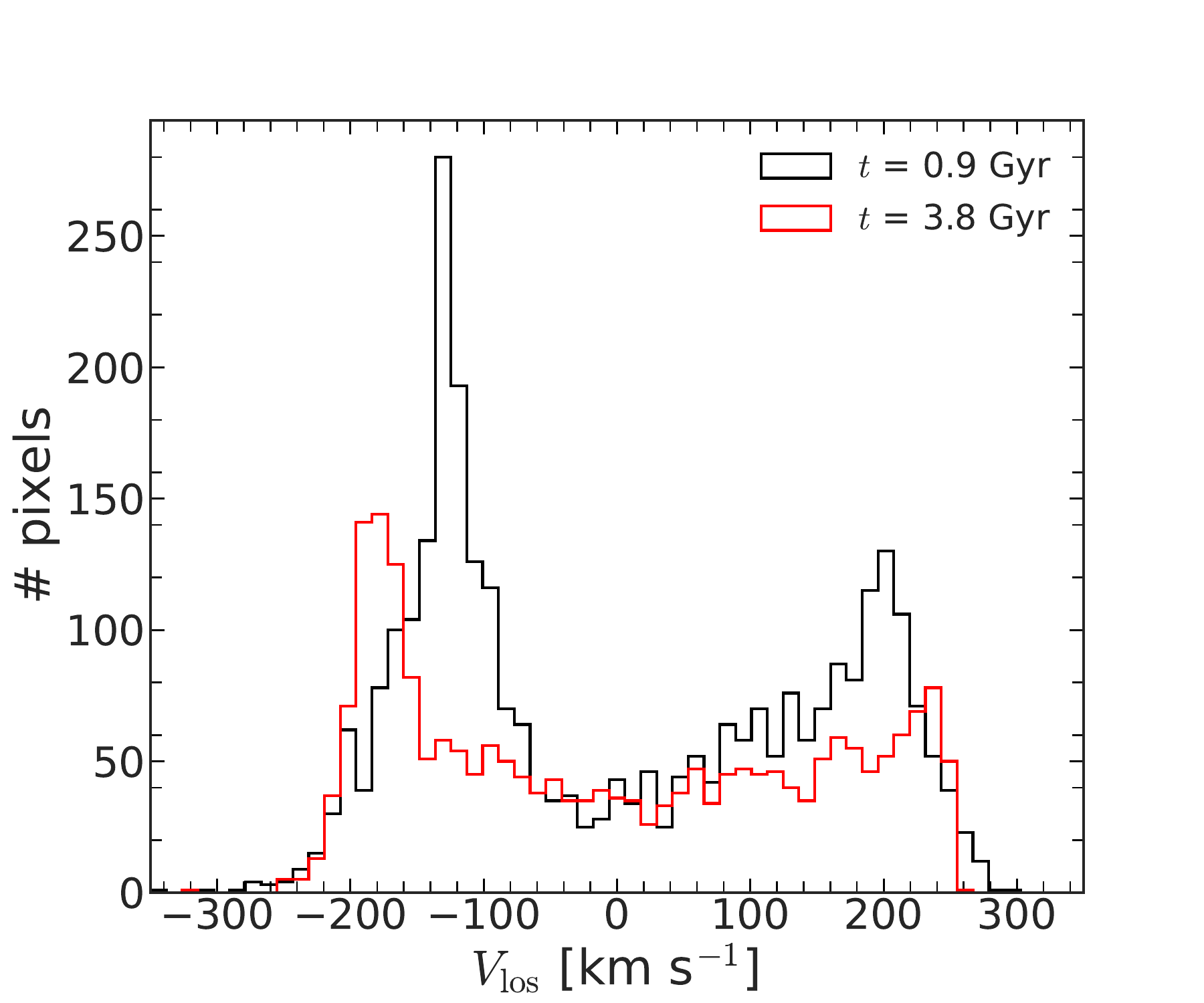}
\caption{Top panels show the gas velocity fields in the edge-on configurations, calculated at two different epochs for the model gSadE001dir33. Bottom panel shows the corresponding distribution of the velocity, calculated separately for the both sides.  At $t = 0.9 \Gyr$, the gas velocity distributions on the either sides are indeed asymmetric.}
\label{fig:lopsided_kinematic_gas}
\end{figure}

Next, we investigate a similar large-scale kinematic asymmetry in the gas velocity field of the minor merger models considered here. First, we construct the gas velocity fields, in edge-on configuration, for the model gSadE001dir33. Two such gas velocity distributions, calculated at $t = 0.9 \Gyr$ and at the end of the simulation run ($t = 3.8 \Gyr$) are shown in Fig.~\ref{fig:lopsided_kinematic_gas} (top panels). Next, we calculate the gas velocity distributions separately for both sides of the host galaxy. The resulting distributions, at $t = 0.9 \Gyr$ and at $t = 3.8 \Gyr$ are shown in Fig.~\ref{fig:lopsided_kinematic_gas} (bottom panels). At $t = 0.9 \Gyr$, the gas velocity distribution shows a high degree of asymmetry when both sides of the host galaxy are considered separately; thereby indicating a presence of kinematic lopsidedness in the gas velocity field as well. This is because, in our models, the stars and the gas are gravitationally-coupled. So, the interstellar gas will also respond in a similar ways as the stars do, to the underlying lopsided potential. We found that this kinematic asymmetry becomes larger after each pericenter passage of the galaxy, similar to what is seen for the stellar velocity fields. However, at the end of the simulation run ($t = 3.8 \Gyr$), the gas velocity distributions in both sides of the host galaxy becomes close to a symmetric distribution. This indicates the absence of a strong kinematic lopsidedness in the gas velocity field, similar to the behaviour at late times for stars (see Fig.~\ref{fig:lopsided_kinematic}).

 Lastly, we probe the persistence of the kinematic lopsidedness in the model gSadE006ret33 where the merger happens at a very late epoch. Following the same methodology, we first construct the distributions of the azimuthal velocity ($v_{\phi}$) in the $x-y$ plane at $t = 0.85 \Gyr$ and $t = 3 \Gyr$. These are shown in Fig.~\ref{fig:lopsided_kinematic} (see bottom panels). A prominent large-scale asymmetry in the velocity distributions exists at $t = 0.85 \Gyr$, similar to other minor merger models considered here; thereby denoting the presence of a strong $m=1$ kinematic lopsidedness. The main difference here is that, the kinematic lopsidedness persists till the end of the simulation run ($t = 3 \Gyr$). We point out that, although an $m=1$ kinematic lopsidedness, excited by a pericenter passage of the satellite galaxy, is short-lived, the repeated pericenter passages in case of the model gSadE006ret33 would drive repeated short-lived kinematic lopsidedness phenomenon.
 
 \section{Comparison with previous works}
\label{sec:comparison_previousWork}
 
 Here, we briefly compare the properties of the $m=1$ lopsidedness (e.g., strength, extent) in the stellar density distribution of the host galaxy in our chosen minor merger models, with the past literature of excitation of lopsidedness via minor mergers or tidal encounters \citep[e.g., ][]{ZaritskyandRix1997,Bournaudetal2005,Mapellietal2008}. We also compare the strength of the $m=1$ lopsided distortion in our models with that from the observed galaxies with lopsidedness, as revealed from the past observational studies.
 \par
 In \citet{ZaritskyandRix1997}, the measured average value of the $m=1$ lopsidedness ($\left<A_1/A_0 \right>$) is $\sim 0.2$ at $R > 1.5 R_{\rm d}$, where the cause of the lopsidedness was conjectured to be tidal interactions. In \citet{Mapellietal2008}, the average values of the $m=1$ lopsidedness ($\left<A_1/A_0 \right>$) is $\sim 0.1$ at $R \sim 2.5 R_{\rm d}$ when the lopsidedness in the stars is excited via fly-by encounter. Similarly, a model presented in \citet[see Fig.~12 there]{Bournaudetal2005}, showed $\left<A_1/A_0 \right> \sim 0.2$ at the initial phase of evolution where the lopsidedness is induced by galaxy interaction and merger. In comparison, the minor merger models considered here, show a stronger $m=1$ lopsidedness ($\left<A_1/A_0 \right> \sim 0.4$ for some models) in the outer disc region ($\sim 4-7 R_{\rm d}$) of the stellar density distribution. In other words, the galaxy interactions involved in these merger models, can generate strong outer disc lopsidedness. However, while comparing these values, it should be borne in mind that the averaging process for measuring $\left<A_1/A_0 \right>$ is carried out over different radial extents for these simulated galaxy models.
 \par
 The location of the occurrence of the prominent $m=1$ lopsidedness also merits some discussion. Past observational studies measured the $m=1$ lopsided distortion up to 3.5 $R_{\rm d}$ where $R_{\rm d}$ is the disc scale-length \citep[e.g., see][]{RixandZaritsky1995,Bournaudetal2005,Zaritskyetal2013} as the near-IR measurements were not available at larger radii due to signal-to-noise constraint. However, the usage of H~{\sc i} as a tracer allowed one to detect the $m=1$ distortion further out in a galaxy. The resulting amplitude of lopsidedness increases with radius up to the optical radius ($\sim 4-5 R_{\rm d}$) and then saturates at larger radii, as measured for the WHISP sample data \citep[e.g., see][]{vanEymeren2011}. While the radial increment of the amplitude of the $m=1$ lopsided distortion is seen in our chosen minor merger models, the $m=1$ lopsided distortion appears predominantly in $4-7 R_{\rm d}$ for our models.
 \par
\begin{figure}
\centering
\includegraphics[width=1.0\linewidth]{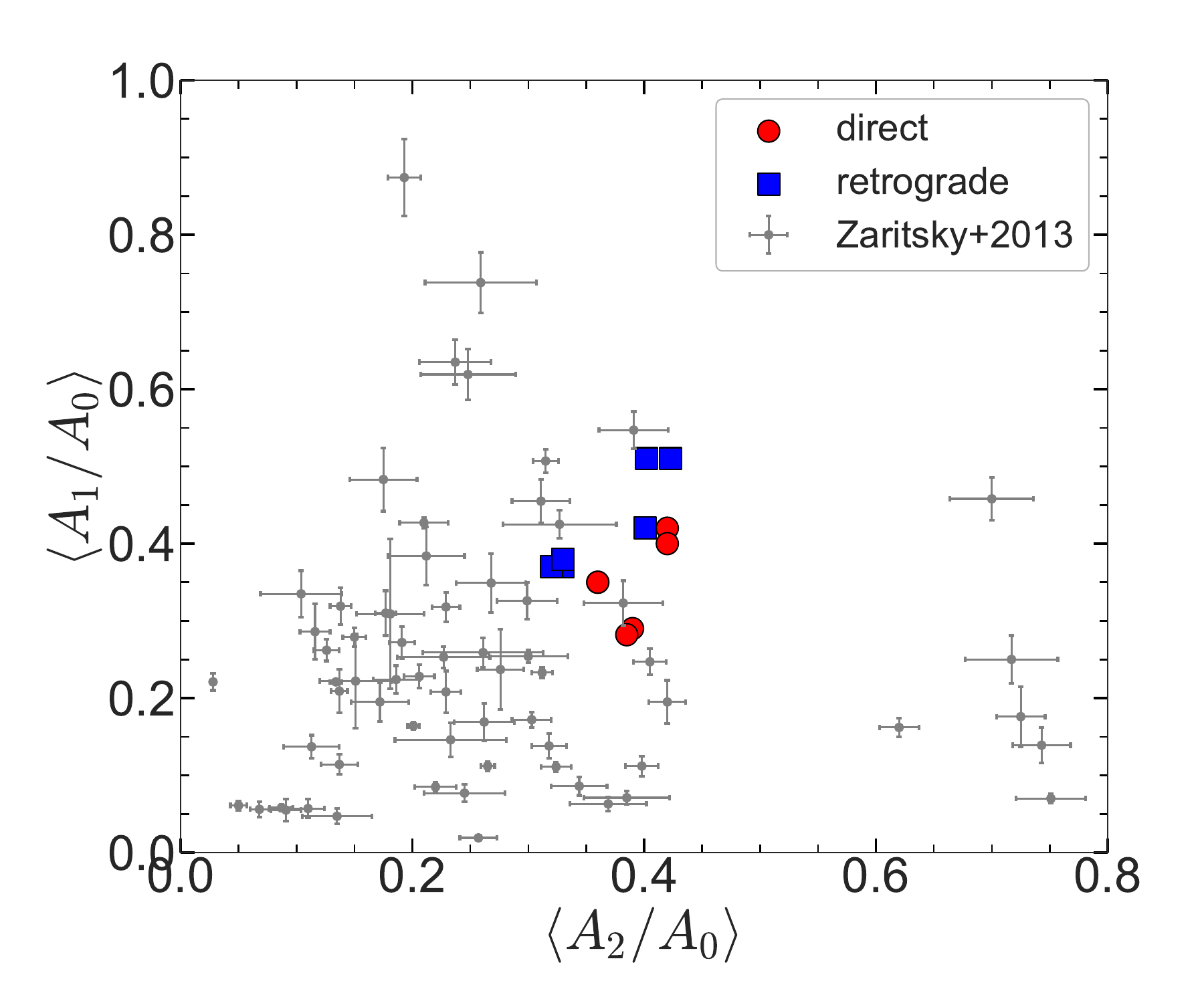}
\caption{Distribution of average values of $m=1$ and $m=2$ Fourier coefficients, $\left<A_1/A_0 \right>$ and $\left<A_2/A_0 \right>$, both calculated after the first pericenter passage, are shown for several gSa-dE0 minor merger models with direct and retrograde orbital configurations. The same are then compared with the measurements by \citep{Zaritskyetal2013} for a sample of galaxies from the S$^4$G catalogue. Here, only galaxies with a bar are selected, for details see text.}
\label{fig:lopsided_comparison_observation}
\end{figure}

Lastly, we compare the strength of the $m=1$ lopsided distortion in our models with that from the observed galaxies with lopsidedness. We note that several models, e.g., gSadE001dir33,  also host a central stellar bar which gets amplified during a pericenter passage of the satellite and subsequently gets weakened in the post-merger remnant \citep[for details, see][]{Ghoshetal2020}. We note that both the bar and the lopsided pattern co-exist after the first pericenter passage of the satellite galaxy. Then, we measure the average values of $m=2$ Fourier coefficient, $\left< A_2/A_0 \right>$, and the average values of the $m=1$ Fourier coefficient $\left< A_1/A_0 \right>$, after the first pericenter passage of the satellite for several minor merger models considered here. The average value of $m=2$ Fourier coefficient is measured in the central region encompassing the bar ($\sim 0.5 - 2 \ R_{\rm d}$) whereas the average value of the $m=1$ Fourier coefficient $\left< A_1/A_0 \right>$ is measured in the outer region encompassing the lopsided pattern ($\sim 4 - 7 \ R_{\rm d}$). This is shown in Fig.~\ref{fig:lopsided_comparison_observation}. Next, to compare with observation, we make use of the measurements of the average values of the $m=1$ and the $m=2$ Fourier coefficients from \citet{Zaritskyetal2013} which provides these measurements  for 163 galaxies selected from the S$^4$G sample \citep{Shethetal2010}. Now, \citet{Zaritskyetal2013} provides the average values of the $m=1$ Fourier coefficient $\left< A_1/A_0 \right>$, measured in two radial extents -- $\left< A_1\right>_i$, measured  in the inner region ($1.5-2.5 R_{\rm d}$), and $\left< A_1\right>_o$, measured in the outer region ($2.5-3.5 R_{\rm d}$). The same is true for the $m=2$ mode as well \citep[for details see description in][]{Zaritskyetal2013}. Therefore, for a uniform comparison with the minor merger models considered here, we have taken the values of  $\left< A_1\right>_o$ and $\left< A_2\right>_i$. Furthermore, we select only those galaxies which host a bar. This is shown in Fig.~\ref{fig:lopsided_comparison_observation}. We note that, although there is a considerable spread in the $\left< A_1\right>_o$ and the $\left< A_2\right>_i$ values for the S$^4$G sample, the measured $\left< A_2/A_0 \right>$ and $\left< A_1/A_0 \right>$ values from our minor merger models tend to lie close to the central clustering of the \citet{Zaritskyetal2013} sample.

\section{Discussion}
\label{sec:discussion}

Here, we discuss a few points regarding this work.\\
\begin{itemize}

\item{We show that a minor merger event can trigger a strong, coherent $m=1$ lopsided pattern in both the density and the velocity distributions of the host galaxy. Thus, minor merger is a plausible avenue to excite lopsidedness in the stellar component of the host galaxies that reside in dense environments (e.g., in groups and in clusters), and in concordance with the finding of previous works \citep[e.g.,][]{Bournaudetal2005,Mapellietal2008}.  Indeed, galaxies residing in groups and clusters, are observationally known to display strong lopsidedness distortion \citep[e.g, see][]{Haynesetal2007}. We show that both direct and retrograde orbital configurations can generate strong, coherent $m=1$ lopsidedness in the host galaxies, as opposed to earlier findings that only retrograde orbital configurations are more favourable to generate lopsided asymmetry \citep{Bournaudetal2005}. However, the longevity of the $m=1$ lopsided pattern indeed depend on the orbital configuration. In our chosen models, for a retrograde orbit, the time of interaction is larger than for a direct orbit with same orbital energy (see Table~\ref{table:key_param}). For example, the lopsidedness is shown to persist for $\sim 2.3 \Gyr$ in the model gSadE006ret33 where the merger happens at a very late epoch. This implies that galaxies experiencing continuous fly-by encounters will show sustained, coherent strong $m=1$ lopsidedness which can be detected in observations. Thus, a minor merger has a significant effect in stirring up the internal dynamics of the host galaxy before the merger is complete. The occurrence of minor merger events are more probable in the early Universe; therefore the secular evolution driven by an $m=1$ lopsided distortion is likely to have a strong influence on the early evolution of galaxies.}

\item{Also, we mention that the satellite merges with the host galaxy, typically around $2 \Gyr$ after the start of the simulation run. As the lopsidedness fades away after the merger happens, therefore, our findings are in apparent tension with the observationally known large number of observed lopsided galaxies in the local Universe as minor mergers are common here \citep[e.g., see][]{Frenketal1988,CarlbergandCouchman1989,LaceyandCole1993,FakhouriandMa2008,Kavirajetal2009}. Furthermore, a galaxy might experience  multiple minor merger events during its entire lifetime  \citep[e.g., see][]{Hopkinsetal2009}. However, we note that in reality, a galaxy can accrete cold gas \citep[e.g.,][]{BirnboimDekel2003,Keresetal2005,DekelBirnboim2006,Ocvriketal2008} either during the merger-phase or at a later stage. Such an asymmetric cold gas accretion can rejuvenate the lopsidedness in the galaxy, as shown in the previous studies \citep[e.g.,][]{Bournaudetal2005,Mapellietal2008}. Further, an isolated galaxy can show signs of an earlier minor merger (e.g., central asymmetric features like an offset disc/bulge, and a tidal stream) even while the outer galaxy seems to have relaxed, as seen in NGC 5523 \citep{Fulmeretal2017}. Thus, while the minor merger event continues to be one plausible mechanism which can excite an $m=1$ lopsidedness in galaxies, as also shown in this work, other mechanisms (e.g., asymmetric gas accretion) are also at play to account for the large abundance of the observed lopsided galaxies.}

\item{As mentioned in section 2, we have considered merger models with  only one inclination angle, namely, $i_1 = 33 \degrees$ as the GalMer library provides models with one inclination for the giant-dwarf minor mergers. We acknowledge that, studying the excitation of lopsidedness in minor merger models with different inclination angles would make this work more complete. Intuitively, the response of the disc would be different for polar ($i_1 =90 \degrees$) and co-planar ($i_1 =0 \degrees$) orbital configurations, and therefore it  will be worth following up in a future study.}

\item{Also, one might wonder whether the low numerical resolution of the GalMer minor merger models used here, could potentially limit the results related to the temporal evolution of the $m=1$ lopsidedness in the host galaxy. During the evolution of disc galaxies, the discrete particle noise indeed affects the evolution of perturbation in galaxies once it is generated \citep[e. g., see][]{VesperiniandWeinberg2000,Choi2007,WeinbergandKatz2007,Weinbergetal2008}. However, we note that the shot noise due to the finite number of particles is not a big issue when the host galaxy is experiencing strong tidal interaction due to a satellite or another perturbing galaxy, as shown in the seminal work by \citet{ToomreandToomre1972}. During such interactions, the change in the potential is overwhelming the change in potential due to the particle noise itself. Furthermore, we checked the excitation and the temporal evolution of the $m=1$ lopsidedness in a dissipationless minor merger merger model where a total of $\sim 2.75 \times 10^7$ particles has been used to model the host galaxy (for details see Appendix~\ref{sec:comparison_with_HighRes}). We found a similar trend of the evolution of the $m=1$ lopsidedness when compared to previously-used GalMer models, namely, a continued pericenter passage excites a strong $m=1$ lopsidedness in the stellar disc. After the merger happens, the $m=1$ lopsidedness gets weakened significantly.  In other words, we do not find any significant difference in the $m=1$ lopsidedness amplitude, maintenance, and the overall temporal evolution as particle number is greatly increased (for details see Appendix~\ref{sec:comparison_with_HighRes}). However, we stress that, in this work we mainly focused on the generation of $m=1$ lopsided perturbation caused directly by the host-satellite interaction. An existing $m=1$  lopsided mode could further be enhanced by a distorted DM halo \citep[e.g.,][]{VesperiniandWeinberg2000,WeinbergandKatz2007} where the distorted halo can drive a long-lasting lopsided feature via resonant interaction. In this scenario, the particle noise is indeed a crucial factor \citep{WeinbergandKatz2007} which can affect/alter the subsequent temporal evolution of the $m=1$ lopsidedness. Also, we caution that although the high-resolution model shows consistent results with the GalMer models (see Appendix~\ref{sec:comparison_with_HighRes}), this high-resolution model is not based on the GalMer models used in earlier sections.
}

\item{We further note that in our models, the DM halo is modelled by a Plummer sphere, i.e., it has a cored density profile at the centre. This choice of a `cored' DM halo density profile is broadly consistent with the observational studies of mass modelling from the observed rotation curve \citep[e.g., see][]{Athanassoulaetal1987,Begemanetal1991,deBloketal2008,deBlok2010,Ohetal2015}. However, we note that past studies of hierarchical galaxy formation in the Lambda cold dark matter paradigm ($\Lambda$CDM) predicted a `cuspy' DM density profile at the centre \citep[e.g., see][]{Navarroetal1996,Navarroetal1997}. For a detailed discussion of the `core-cusp' problem, the reader is referred to \citet{deBlok2010}. It might be interesting to investigate the  excitation and the evolution of an $m=1$ lopsided distortion in a cuspy DM halo; however, it is beyond the scope of the present paper. Furthermore, we note that, stellar and DM halo masses of the gSa-type host galaxies in our selected minor merger models, are similar. This could be a strong assumption when one wants to study the DM-baryonic coupling of the perturbations. However, we point out that the baryonic fraction in our gSa-type galaxy models is similar to the baryonic content, within the optical radius, for the typical high-surface-brightness galaxies \citep[e.g., see][]{deBloketal2001}. Furthermore, we mention that for the GalMer model, the DM halo is truncated, and it is not simulated up to the Virial radius. This is justified since we begin the GalMer merger simulations, when the two galaxies are close enough, so that they have already merged their outer haloes (say $200 \kpc$ sizes), and are embedded in a common quasi-spherical DM envelope. Thus, this might not have strong influence on the dynamics of the $10- \kpc$ scale merging, and the $m=1$ lopsidedness.}

\item{Lastly, we mention that a disc-DM halo off-set configuration was previously shown to lead to a strong central lopsidedness in the central few kpc region \citep[e.g, see][]{PrasadandJog2017} as the central disc mass dominates the halo so that the off-set halo acts as a perturbation on the disc. Although such a disc-DM halo off-set exists (for a short time-scale) in our minor merger models, we do not find any strong lopsidedness in the central disc regions of the host galaxy.}
\end{itemize}

\section{Conclusion}
\label{sec:conclusion}

In summary, we investigated the dynamical impact of minor merger of galaxies (mass ratio 1:10) on the excitation of an $m=1$ lopsided distortion in the stellar density and the velocity fields. We also studied the generation of a stellar disc-DM halo off-set configuration in the host galaxy during a minor merger event. We selected a set of minor merger models, with varying orbital energy, orientation of orbital spin vector, morphology of host galaxy from the GalMer library of galaxy merger simulation. Our main findings are :\\

\begin{itemize}

\item{A minor merger event can trigger a prominent $m=1$ lopsided distortion in the stellar density distribution of the host galaxy. The strength of the lopsided distortion undergoes a transient amplification phase after each pericenter passage of the satellite. However, the lopsidedness fades away after the merger happens and the post-merger remnant gets time ($\sim 500-850 \Myr$) to readjust itself.  This broad trend holds true for a wide range of orbital configurations considered here. In addition, a \textit{delayed} minor merger can drive a prolonged ($\sim 2-2.5 \Gyr$) lopsidedness due to continued pericenter passages of the satellite.}

\item{The $m=1$ lopsided pattern is shown to rotate in the disc with a well-defined pattern speed. The pattern speed of the $m=1$ lopsidedness is smaller than the pattern speed of the $m=2$ bar, when measured simultaneously at a same epoch. Moreover, the $m=1$ lopsided distortion rotates in a retrograde sense with respect to the $m=2$ bar mode. This gives rise to a dynamical scenario of a bar-lopsidedness resonance overlap.}

\item{The stellar and the gas velocity fields of the host galaxy also displays a large-scale kinematic lopsidedness after each pericenter passage of the satellite galaxy. The temporal evolution of the morphological and the kinematic lopsidedness closely follow each other. }

\item{An interaction with a satellite galaxy also excites an off-set between the stellar disc and the DM halo of the host galaxy. The resulting off-set is 2-3 times of the softening length of the simulation. This off-set is rather short-lived, and is most prominent after each pericenter passage of the satellite. This holds true for a wide range of orbital configurations considered here.}

\end{itemize}

\section*{Acknowledgement}

We thank the anonymous referee for detailed constructive comments that helped to improve our paper. The authors acknowledge support from an Indo-French CEFIPRA project (Project No.: 5804-1). 
CJJ thanks the DST, Government of India for support via a J.C. Bose fellowship (SB/S2/JCB-31/2014).
This work makes use of the publicly available GalMer library of galaxy merger simulations which is a part of {\sc HORIZON} project (\href{http://www.projet-horizon.fr/rubrique3.html} {http://www.projet-horizon.fr/rubrique3.html}).

\section*{Data availability}

The simulation data of minor merger models used here is publicly available from the URL  \href {http:/ /galmer.obspm.fr}{http:/ /galmer.obspm.fr}. The measurements of average values of $m=1$  lopsided distortion and $m=2$ bar mode for the sample of S$^4$G galaxies are publicly available from  \href {https://doi.org//10.26093/cds/vizier.17720135}{this URL}. The newly-added high-resolution simulation is taken from the paper \citet{Baptisteetal2017}.

\bibliography{my_ref}{}
\bibliographystyle{mnras}


\appendix

\section{Evolution in isolation}
\label{sec:isolatedEvolution}
Earlier we showed the generation of an off-set between the stellar and the DM halo centres and the excitation of a prominent $m=1$ lopsided distortion in the density and the velocity fields of the host galaxy in the minor merger models considered here. 
 However, in order to demonstrate conclusively that these physical phenomena are indeed triggered by a minor merger event, one needs to study the evolution of the host galaxy model in isolation.

\begin{figure}
\centering
\includegraphics[width=1.0\linewidth]{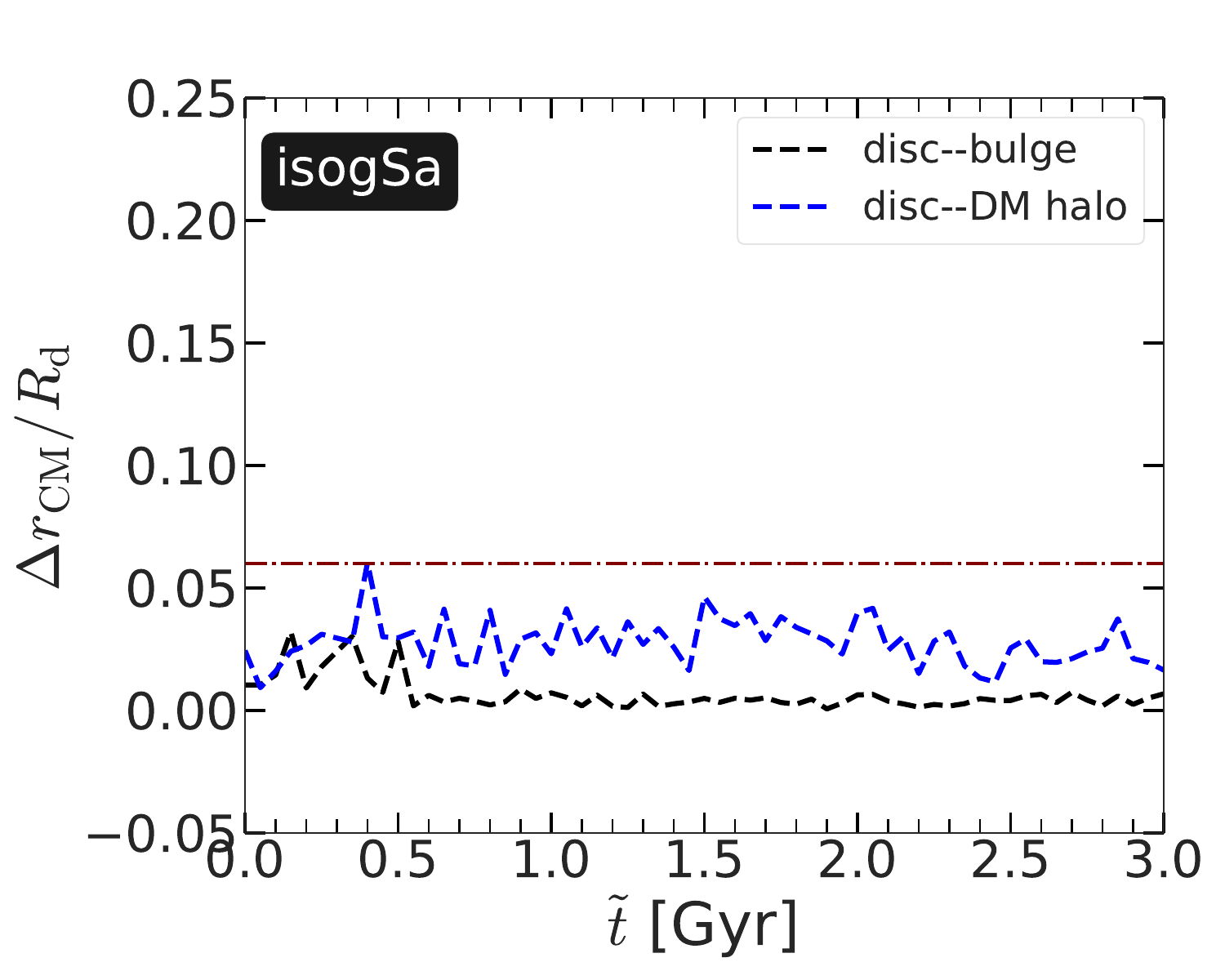}
\medskip
\includegraphics[width=1.0\linewidth]{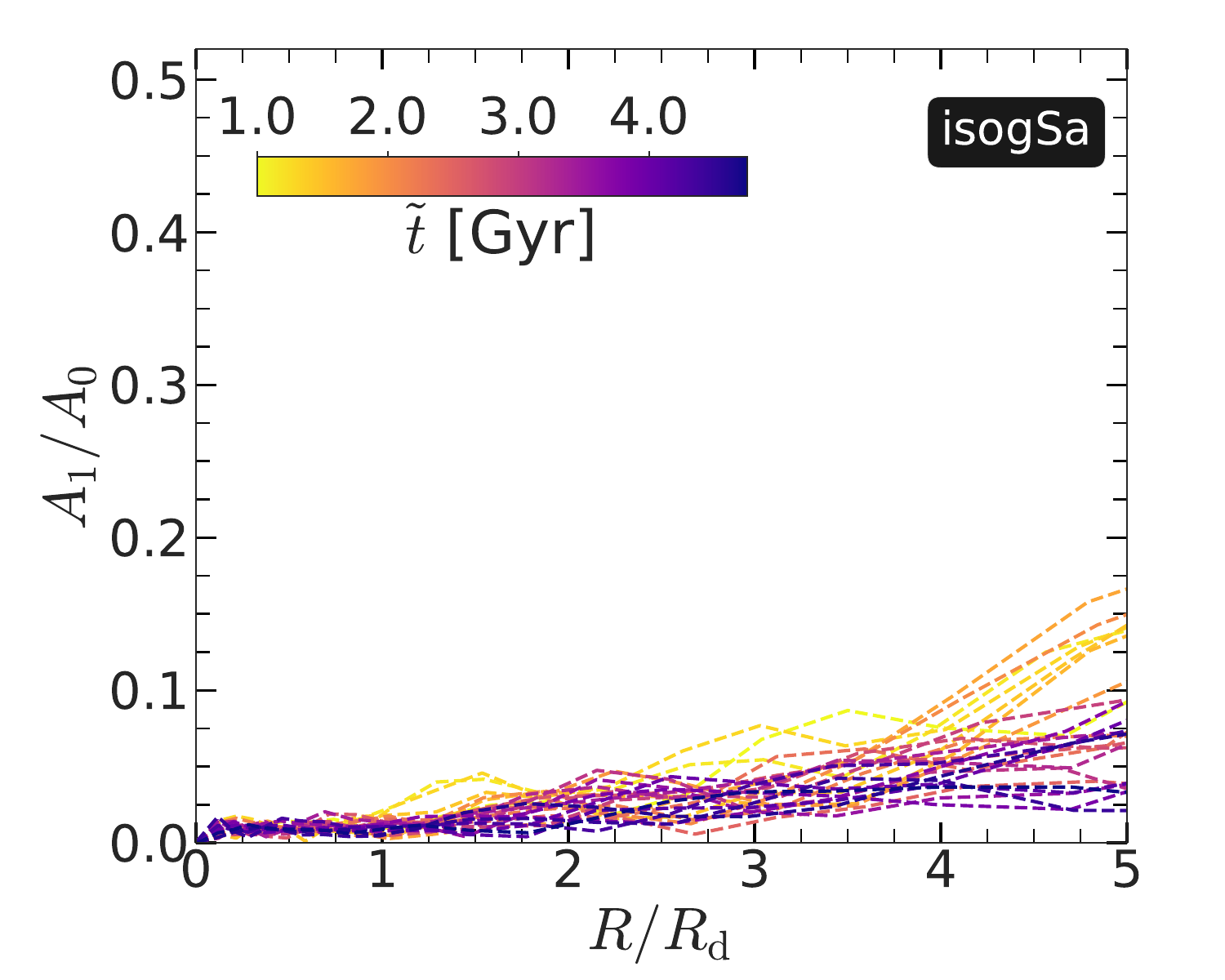}
\caption{Evolution in isolation -- top panel shows the distances between the disc-bulge and the disc-DM halo centres, as a function of time for the isolated isogSa model. The horizontal dash line (in maroon) denotes the softening length ($\epsilon = 200 \pc$). Bottom panel shows the corresponding radial profiles of $m=1$ Fourier coefficients as a function of time. Here, $\tilde t - t = 1 \Gyr$, for details see text.}
\label{fig:isomodel}
\end{figure}

To address this, we study the model for the host galaxy of gSa-type (isogSa) in isolation for $4.9 \Gyr$. We point out that the models are evolved in isolation for $1 \Gyr$ before the merger simulation sets in (see Section~\ref{sec:simu_setup}). Consequently, there is a time delay of $1 \Gyr$ between the isolated and the minor merger models, i.e., $\tilde t - t = 1 \Gyr$. Fig.~\ref{fig:isomodel} (top panel) shows the corresponding time evolution of the distance between the density-weighted centres of the disc and the DM halo. As seen clearly, no such prominent off-set between the stars and the DM halo is created when it is evolved in isolation. The separation between these two centres always remains below the the softening length used in the simulation. 
\par
Fig.~\ref{fig:isomodel} (bottom panel) also shows the radial profiles of the $m=1$ Fourier coefficients as a function of time.   A smaller value of the  Fourier coefficient ($A_1/A_0$), when compared with a minor merger model, e.g., gSadE001dir33 (compare with Fig.~\ref{fig:lopsided_radial}), demonstrates the absence of a coherent lopsided pattern in the stellar density distribution. We mention that for some time-steps, a small fraction of stellar particles remained scattered in the very outer region ($\ge 4.5 R_{\rm d}$), thus giving a spuriously high value of the $m=1$ Fourier coefficient in that radial range. Those values should be disregarded as a signature of coherent $m=1$ lopsided distortions. We also checked that unlike a minor merger model, an evolution in isolation does not excite a kinematic lopsidedness in the stellar velocity field as well, for brevity we have not shown here. Thus,  the generation of an off-centred stellar disc-DM halo configuration as well as the excitation of a prominent $m=1$ lopsided distortion in both the stellar density and velocity fields of the host galaxy can indeed be attributed to the dynamical effects of a minor merger event.

\section{Calculation of intrinsic circular velocity}
\label{sec:circvel}

Here, we briefly describe how the circular velocity ($v_c$) is calculated from the intrinsic particle distribution. First, at any given time-step, we calculate the radial profiles azimuthal velocity ($v_{\phi}$) and the associated velocity dispersion components along the radial and the azimuthal directions ($\sigma_{R}$, $\sigma_{\phi}$), respectively. The circular velocity ($v_{\rm c}$) can be calculated from these quantities while correcting for the asymmetric drift via the equation \citep{BiineyTremaine2008}
\begin{equation}
v_{\rm c}^2 = v_{\phi}^2+ \sigma_{\phi}^2 -\sigma_{R}^2 \left(1 + \frac{\rm d \ ln \ \rho}{\rm d \ ln \  R} + \frac{\rm d \ ln \ \sigma^2_R}{\rm d \ ln \  R}\right)\,.
\label{eq:asy_drift1}
\end{equation}

\begin{figure}
\centering
\includegraphics[width=1.0\linewidth]{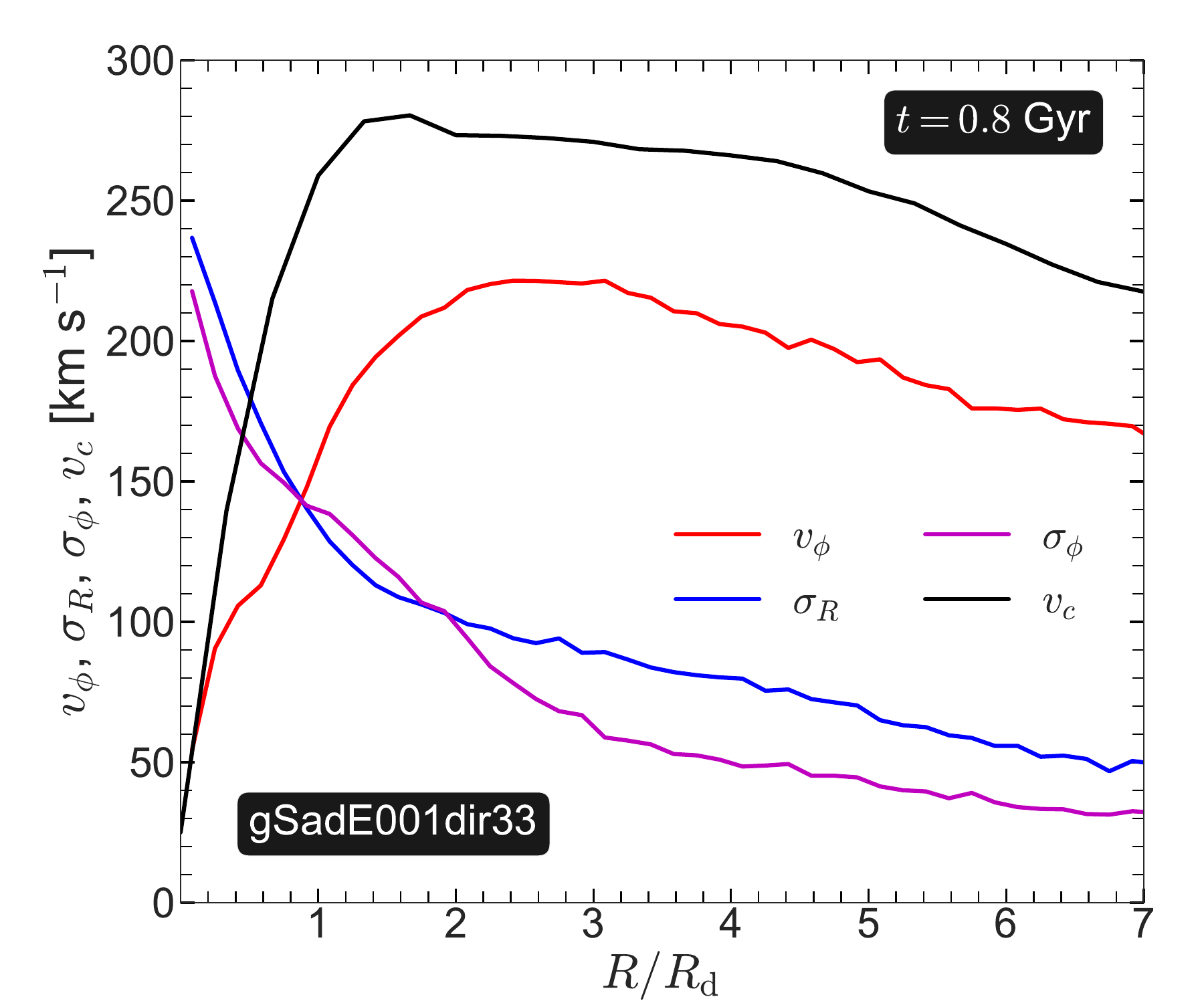}
\caption{Radial variation of the intrinsic circular velocity, $v_c$ (calculated using Eq.~\ref{eq:asy_drift1}), together with the radial variations of the azimuthal velocity ($v_{\phi}$), the radial and azimuthal velocity dispersion components ($\sigma_{R}$, $\sigma_{\phi}$) are shown at $t = 0.8 \Gyr$ for the model gSadE001dir33.}
\label{fig:circvel_demo}
\end{figure}

The resulting radial profile of circular velocity ($v_c$), along with the radial profiles of $v_{\phi}$, $\sigma_{R}$, and $\sigma_{\phi}$, calculated at $t= 0.8 \Gyr$ for the model gSadE001dir33 are shown in Fig.~\ref{fig:circvel_demo}.

\section{Comparison with a high-resolution merger model}
\label{sec:comparison_with_HighRes}

\begin{figure*}
\centering
\includegraphics[width=1.0\linewidth]{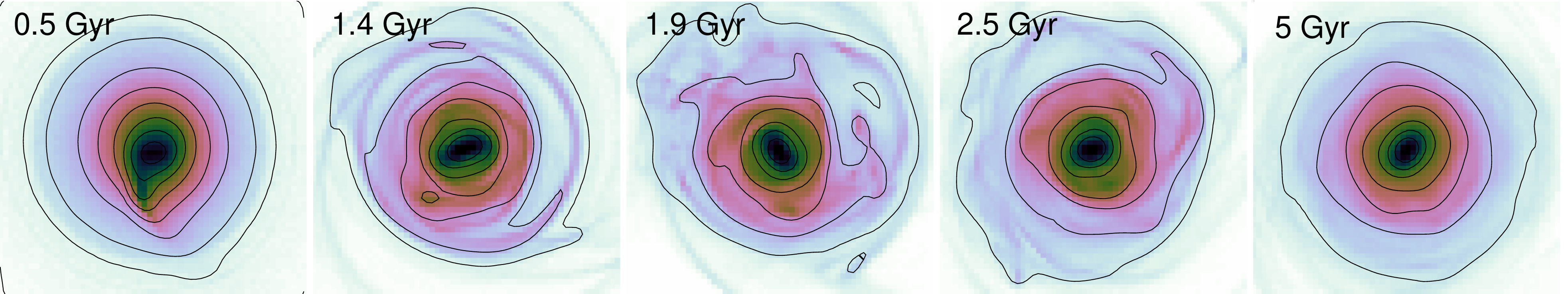}
\begin{multicols}{2}
\includegraphics[width=1.\linewidth]{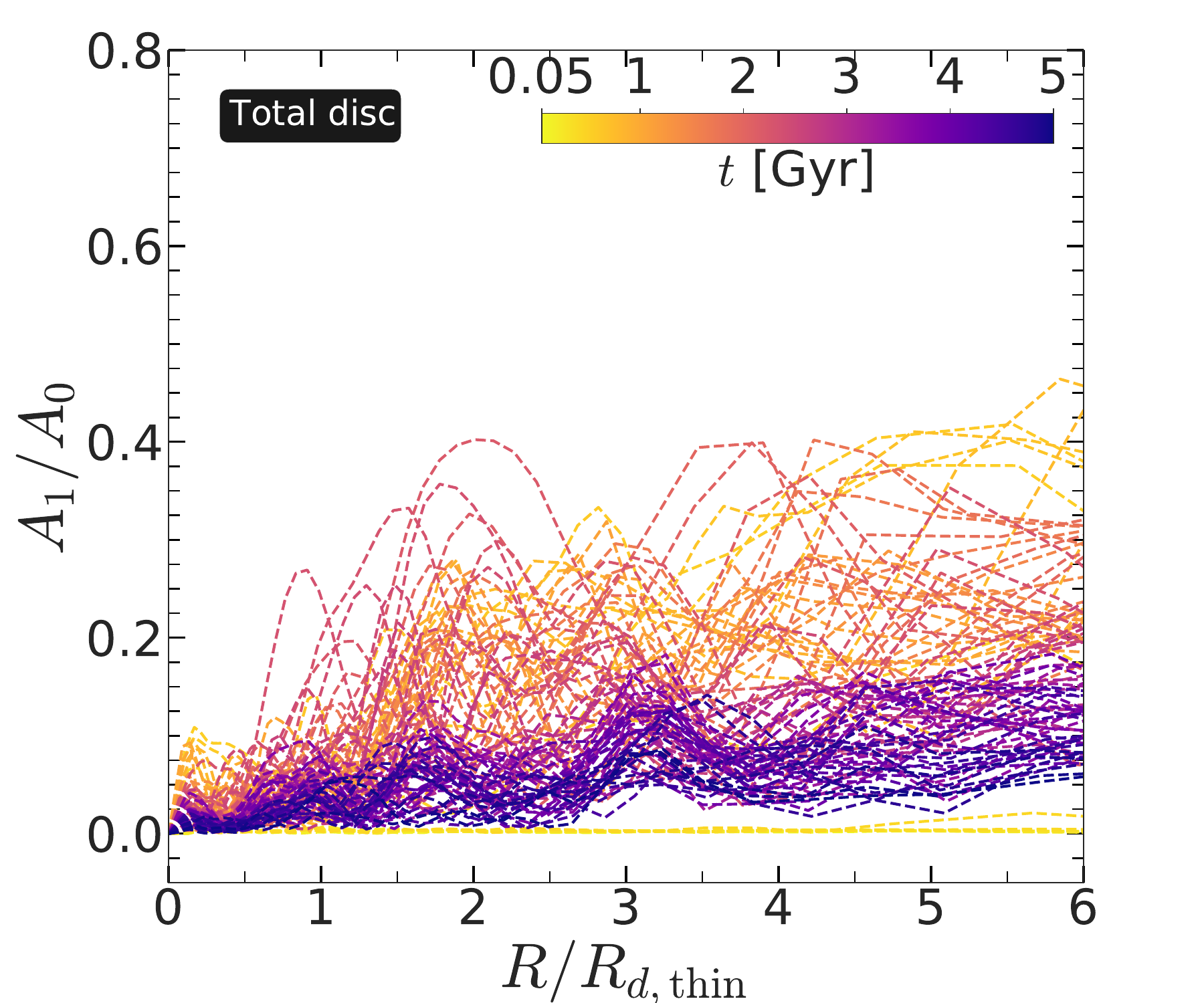}\par
\includegraphics[width=1.\linewidth]{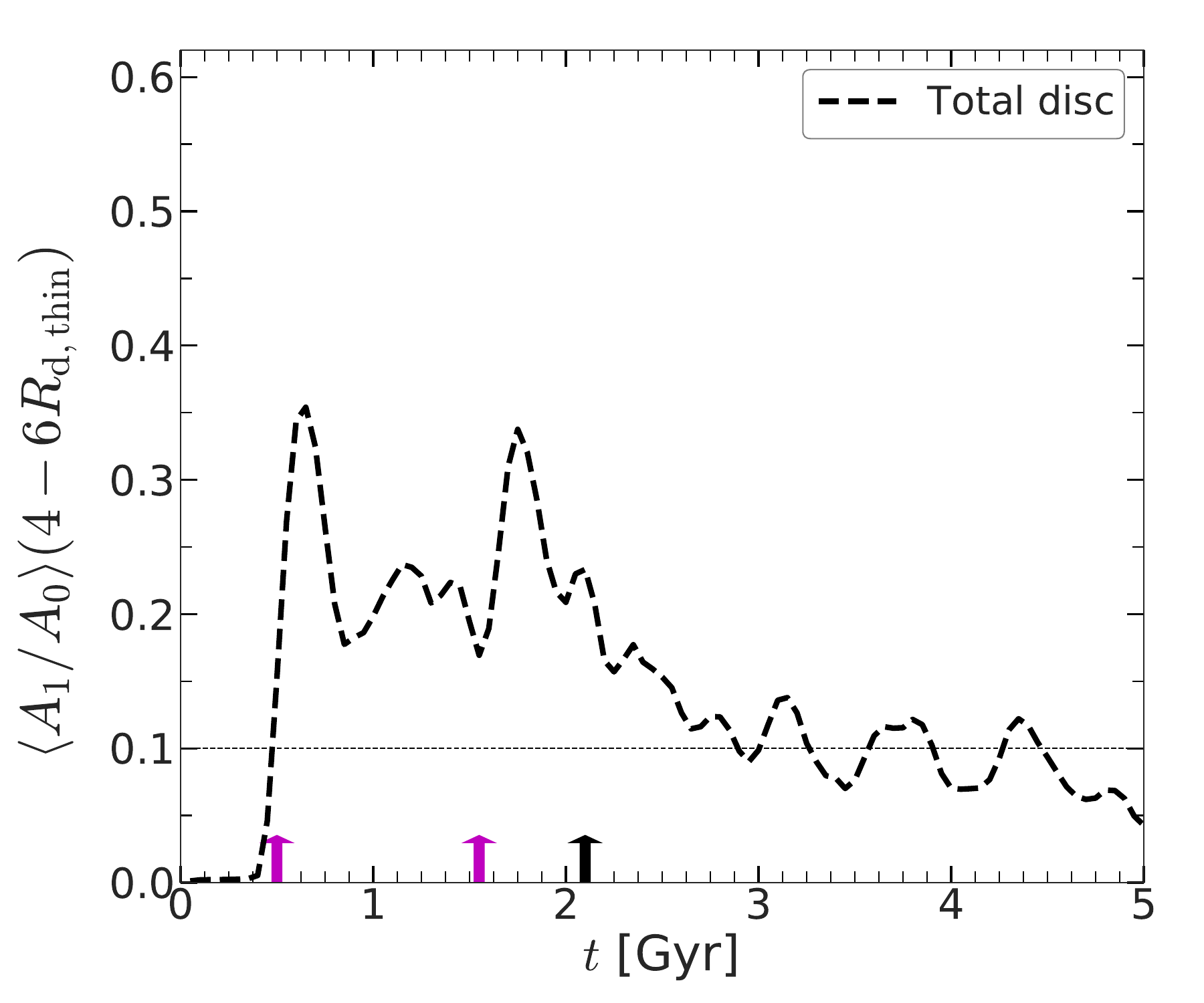}\par
\end{multicols}
\caption{High-Res model: top panels show the  face-on density maps of the total (thin+intermediate+thick) disc particles of the host galaxy at different epochs before and after the merger happens. Black solid lines denote the contours of constant density. Bottom left panel shows the radial profiles of the $m=1$ Fourier coefficient ($A_1/A_0$) at different time-steps. Simulation run-time is shown in the colour bar. Bottom right panel shows the evolution of the median value of the $m=1$ Fourier coefficient $A_1/A_0$, calculated within the radial extent of $4-6 R_{\rm d, thin}$, as a function of time. The black horizontal line denotes $A_1/A_0=0.1$, and is used as a demarcation for the onset of the $m=1$ lopsidedness. Magenta vertical arrows denote the epochs of the pericenter passages while black vertical arrow denotes the epoch of the merger with the satellite galaxy. Here, $R_{\rm d, thin} = 4.7 \kpc$.}
\label{fig:compare_highres}
\end{figure*}
Here, we present the results related to the excitation of an $m=1$ lopsidedness and its temporal evolution for a dissipationless minor merger simulation which uses a higher number of particles as compared to the previously-used GalMer models. This model is referred to as `High-Res' model, and it is taken from \citet{Baptisteetal2017}. For the sake of completeness, below we provide a brief description about the details of the simulation set-up for this High-Res model (see section~\ref{sec:Appen_simSetup_highresModel}). The corresponding generation and the temporal evolution of an $m=1$ lopsidedness in presented in section~\ref{sec:Appen_m1Evolution_highresmodel}.

\subsection{Simulation set-up : High-Res model}
\label{sec:Appen_simSetup_highresModel}

This model is used here to investigate whether the generic trends about the excitation and the temporal evolution of an $m=1$ lopsidedness depend on the particle resolution of a minor merger model.
The host galaxy is comprised of a thin disc, an intermediate disc, and a thick disc component which are embedded in a DM halo. Each of the stellar disc components is modelled with a Miyamoto-Nagai density profile whereas the DM halo density profile is modelled with a Plummer sphere. For details of the scale lengths and heights of these density profiles, the reader is referred to \citet[see Table 1 there]{Baptisteetal2017}. The host galaxy also contains a number of globular clusters which are treated as point masses. The total number of particles ($N_{\rm tot}$) used for modelling the host galaxy is $2.75 \times 10^7$. The number of particles used for the thin disc ($N_{\rm thin}$), the intermediate disc ($N_{\rm inter}$), and for the thick disc ($N_{\rm thick}$) are $1 \times 10^7$, $6 \times 10^6$, and $4 \times 10^6$, respectively whereas the number of particles used for the DM halo ($N_{\rm DM}$) is $5 \times 10^6$. The masses of the thin, intermediate, and the thick discs are $2.55 \times 10^{10} M_{\sun}$, $1.55 \times 10^{10} M_{\sun}$, and $1 \times 10^{10} M_{\sun}$, respectively \citep[for details see Table 1 in][]{Baptisteetal2017}. The satellite galaxy is the re-scaled version of the host galaxy, with its mass and total number of particles are one-tenth of the host galaxy while its size is reduced by a factor $\sqrt{10}$ \citep{Baptisteetal2017}.
\par
 In a reference frame whose  $x-y$ plane coincides with that of the host galaxy, and the $z$-axis oriented along the spin of the host galaxy, the orbital plane configuration is completely defined by $\theta_{\rm orb}$, and $\phi_{\rm orb}$. The angle $\theta_{\rm orb}$ denotes the angle between the intersection of the orbital plane with the $x-y$-plane and the $x$-axis whereas the angle  $\phi_{\rm orb}$ denotes  the angle between the intersection of the orbital plane with the $x-z$-plane and the $x$-axis. Similarly, the satellite's orientation is defined by $\theta_{\rm sat}$, and $\phi_{\rm sat}$. Initially, the satellite is placed at a distance of $100 \kpc$ from the host galaxy, and then placed in a direct orbit. For the initial velocities, angular momenta, and other orbital configuration parameters, the reader is referred to \citet[see Table 2 there]{Baptisteetal2017} The epochs of the first and the second pericenter passages of the satellite are $t = 0.5 \Gyr$, and $t = 1.55 \Gyr$, respectively while the satellite merges with the host galaxy at $t = 2.1 \Gyr$ \citep[see Fig. 1 there]{Baptisteetal2017}. The simulation is run for $5 \Gyr$.

\subsection{Excitation and temporal evolution of m=1 lopsidedness : High-Res model}
\label{sec:Appen_m1Evolution_highresmodel}

Fig.~\ref{fig:compare_highres} shows the density distribution of the total (thin+intermediate+thick) disc particles in the face-on configuration, before and the merger happens. Soon after the first pericenter passage ($T_{1, peri} = 0.5 \Gyr$), a prominent $m=1$ asymmetry in the density distribution appears. Similar signature of a global $m=1$ density asymmetry is also seen after the second pericenter passage ($T_{2, peri} = 1.55 \Gyr$). However, at the end of the simulation run ($t = 5 \Gyr$), the corresponding density distribution becomes more axisymmetric. To quantify that, we calculated the radial profiles of the $m=1$ Fourier coefficients (using all disc particles of the host galaxy) at different time-steps. This is shown is Fig.~\ref{fig:compare_highres} (see bottom left panel). Initially, the $A_1/A_0$ values are close to 0, thereby denoting the absence of a prominent $m=1$ lopsidedness in the stellar density distribution. After, the first pericenter passage of the satellite galaxy happens, the radial profiles of the $A_1/A_0$ show higher non-zero value, thereby demonstrating the presence of a prominent $m=1$ lopsidedness. We note that, in this High-Res model, a one-armed spiral is also triggered (similar to the GalMer models) in the inner region ($R \leq 4 R_{\rm d, thin}$, where $R_{\rm d, thin} = 4.7 \kpc$) which in turn, gives rise to a hump-like feature in the radial profiles of $A_1/A_0$. At the end of the simulation run ($t = 5 \Gyr$), the radial profiles of $A_1/A_0$ again show a lesser value ($\sim 0.1-0.15$), thereby implying that the $m=1$ lopsidedness has eventually faded away. Lastly, to quantify the temporal evolution of the $m=1$ lopsided distortion, we calculated the median values of $A_1/A_0$ in the radial extent $4 \leq R/R_{\rm d, thin} \leq 6$ as a function of time. This is shown in Fig.~\ref{fig:compare_highres} (see bottom right panel). As seen clearly, a prominent $m=1$ lopsidedness appears only after the first pericenter passage of the satellite, then it decays with time, and reappears after the second pericenter passage of the satellite. However, after the satellite galaxy merges with the host galaxy, the corresponding median value of $A_1/A_0$ in the chosen radial extent ($\avg{A_1/A_0}$) starts decreasing significantly. Around $t = 3.5 \Gyr$, the corresponding value of $\avg{A_1/A_0}$ starts oscillating around $\avg{A_1/A_0} =0.1$, thereby implying the $m=1$ density lopsidedness has decayed substantially. At the end of the simulation run ($t = 5 \Gyr$), the value of $\avg{A_1/A_0}$ lies below 0.1, thereby accentuating that the $m=1$ density lopsidedness is faded away. Therefore, to summarise, the High-res model shows a similar trend in the excitation and the temporal evolution of the $m=1$ density lopsidedness when compared with the previously-shown GalMer model having lower particle resolution.

\bsp
\label{lastpage}

\end{document}